\documentclass[ALICE,manyauthors]{cernphprep}

\pdfoutput=1
\usepackage{graphicx}
\usepackage{bm}      
\usepackage{amssymb} 
\usepackage{amsfonts}
\usepackage{graphics}
\usepackage{epsfig}
\usepackage[normalem]{ulem}
\usepackage[usenames]{color}
\usepackage{multirow}
\usepackage{units}
\usepackage{booktabs}
\usepackage{tabularx}
\usepackage{hyperref}

\newcommand{\ZNA}          {\rm{ZNA}}
\newcommand{\ZNC}          {\rm{ZNC}}

\newcommand{\VZEROA}       {\rm{VZERO-A}}
\newcommand{\VZEROC}       {\rm{VZERO-C}}
\newcommand{\pip}          {$\pi^{+}$}
\newcommand{\pim}          {$\pi^{-}$}
\newcommand{\kap}          {K$^{+}$}
\newcommand{\kam}          {K$^{-}$}
\newcommand{\pbar}         {$\rm\overline{p}$}
\newcommand{\kzero}        {\ensuremath{{\rm K}^{0}_{S}}}
\newcommand{\vzero}        {\ensuremath{{\rm V}^0}}
\newcommand{\lmb}          {\ensuremath{\Lambda}}
\newcommand{\almb}         {\ensuremath{\bar{\Lambda}}}
\newcommand{\allpart}      {$\pi^{\pm}$, K$^{\pm}$, \kzero, p(\pbar) and \lmb(\almb)}
\newcommand{\allpi}        {$\pi^{\pm}$}
\newcommand{\allk}         {K$^{\pm}$}
\newcommand{\allp}         {p(\pbar)}

\newcommand{\dedx}         {\ensuremath{\mathrm{d}E/\mathrm{d}x}}

\newcommand{\pp}           {pp}

\newcommand{\PbPb}         {\mbox{Pb--Pb}}
\newcommand{\pPb}          {\mbox{p--Pb}}

\newcommand{\dNdeta}       {\ensuremath{\mathrm{d}N_\mathrm{ch}/\mathrm{d}\eta}}

\newcommand{\s}            {\ensuremath{\sqrt{s}}}
\newcommand{\pt}           {\ensuremath{p_{\rm T}}}
\newcommand{\hlab}         {\ensuremath{\eta_{\rm lab}}}
\newcommand{\ynn}         {\ensuremath{y_{\rm NN}}}
\newcommand{\ycms}         {\ensuremath{y_{\rm CMS}}}

\newcommand{\ppi}          {\ensuremath{{\rm p}/\pi}}
\newcommand{\kpi}          {\ensuremath{{\rm K}/\pi}}
\newcommand{\lpi}          {\ensuremath{{\rm \Lambda}/\pi}}
\newcommand{\mt}           {\ensuremath{m_{\rm T}}}
\newcommand{\snn}          {\ensuremath{\sqrt{s_{\rm NN}}}}
\newcommand{\snnbf}        {\ensuremath{\mathbf{{\sqrt{s_{\mathbf NN}}}}}}

\newcommand{\Tfo}          {\ensuremath{{T}_{\rm kin}}}
\newcommand{\Tch}          {\ensuremath{{T}_{\rm ch}}}

\newcommand{\avbT}         {\ensuremath{\left< \beta_{\rm T}\right>}}
\newcommand{\avpT}         {\ensuremath{\left< \pt \right>}}
\newcommand{\muB}          {\ensuremath{\mu_{B}}}

\newcommand{\gevc}         {\ensuremath{{\rm GeV}/c}}
\newcommand{\mevc}         {\ensuremath{{\rm MeV}/c}}
\newcommand{\avg}[1]       {\ensuremath{\left\langle#1\right\rangle}}

\graphicspath{{./img/}}

\begin{document}
\begin{titlepage}
\PHnumber{2013-135}
\PHdate{July 25, 2013}
\title{Multiplicity Dependence of Pion, Kaon, Proton and Lambda Production in p--Pb Collisions at \snnbf~=~5.02~TeV}
\ShortTitle{Multiplicity Dependence of \allpart\ in p--Pb Collisions}
\Collaboration{ALICE Collaboration
         \thanks{See Appendix~\ref{app:collab} for the list of collaboration
                      members}}
\ShortAuthor{ALICE Collaboration}
\begin{abstract}
In this Letter, comprehensive results on \allpart\ production at
mid-rapidity ($0 < \ycms < 0.5$) in \pPb\ collisions at
\snn~=~5.02~TeV, measured by the ALICE detector at the LHC, are
reported.  The transverse momentum distributions exhibit a hardening as a 
function of event multiplicity, which is stronger for heavier particles. 
This behavior is similar to what has been observed in pp and Pb--Pb collisions at the LHC.
The measured \pt\ distributions are compared to d--Au, Au--Au and
 Pb--Pb results at lower energy and with predictions based on 
QCD-inspired and hydrodynamic models.

\end{abstract}
\end{titlepage}
\newif\ifplb
\plbtrue

\section{Introduction}

High-energy heavy-ion (AA) collisions offer a unique possibility to study
nuclear matter under extreme conditions, in particular the deconfined
quark-gluon plasma which has been predicted by quantum chromodynamics
(QCD)~\cite{Cabibbo:1975ig, Shuryak:1978ij, McLerran:1980pk,
  Laermann:2003cv}.  The interpretation of heavy-ion results depends
crucially on the comparison with results from smaller collision
systems such as proton-proton (\pp) or proton-nucleus (pA).

The bulk matter created in high-energy nuclear reactions can be
quantitatively described in terms of hydrodynamic and statistical
models. The initial hot and dense partonic matter rapidly expands and
cools down, ultimately undergoing a transition to a hadron gas
phase~\cite{Muller:2006ee}.  The observed ratios of particle
abundances can be described in terms of statistical
models~\cite{Andronic:2008gu,Cleymans:2006xj}, which are governed
mainly by two parameters, the chemical freeze-out temperature \Tch\
and the baryochemical potential \muB\, which describes the net baryon
content of the system.
These models provide an accurate description of the data over a large
range of center-of-mass energies~(see
e.g.~\cite{BraunMunzinger:2003zz}), but a surprisingly large deviation
(about 50\%) was found for the proton production yield at the LHC~\cite{prl-spectra,
  Abelev:2013vea}.
During the expansion phase, collective hydrodynamic flow develops from
the initially generated pressure gradients in the strongly interacting
system. This results in a characteristic dependence of the shape of
the transverse momentum (\pt) distribution on the particle mass, which
can be described with a common kinetic freeze-out temperature parameter \Tfo\
and a collective average expansion velocity
\avbT~\cite{Schnedermann:1993ws}.

Proton-nucleus (pA) collisions are intermediate between
proton-proton (pp) and nucleus-nucleus (AA) collisions in terms of
system size and number of produced particles. Comparing particle
production in pp, pA, and AA reactions has frequently been used to
separate initial state effects, linked to the use of nuclear
beams or targets, from final state effects, linked to the presence of hot and
dense matter. At the LHC, however, the pseudorapidity density of final state
particles in pA collisions reaches values which can become
comparable to semi-peripheral Au--Au ($\sim$60\% most central) and Cu--Cu ($\sim$30\% most central) collisions at top RHIC energy~\cite{Alver:2010ck}.
Therefore the assumption that final state dense matter effects can be
neglected in pA may no longer be valid.  In addition, pA collisions allow for the
investigation of fundamental properties of QCD: the relevant part of
the initial state nuclear wave function extends to very low fractional
parton momentum $x$ and very high gluon densities, where parton
shadowing and novel phenomena like saturation, e.g. as implemented in
the Color Glass Condensate model (CGC), may become
apparent~\cite{McLerran:1993ni, Gelis:2010nm}.

Recently, measurements at the LHC in high multiplicity pp and \pPb\
collisions have revealed a near-side long-range ``ridge'' structure in
the two-particle
correlations~\cite{Khachatryan:2010gv,CMS:2012qk}.
The observation of an unexpected ``double-ridge'' structure in the two-particle correlations in high-multiplicity \pPb\ collisions has also been reported~\cite{Abelev:1497210,Aad:2012gla,Aad:2013fja,Chatrchyan:2013nka}. 
This is flat and long-range in pseudo-rapidity $\Delta\eta$ and
modulated in azimuth approximately like $\cos(2 \Delta\phi)$, where
$\Delta\eta$ and $\Delta\phi$ are the differences in pseudo-rapidity
$\eta$ and azimuthal angle $\phi$ between the two particles. Various
mechanisms have been proposed to explain the origin of this
double-ridge like structure. Both a CGC
description~\cite{Dusling:2013oia}, based on initial state nonlinear
gluon interactions, as well as a model based on hydrodynamic
flow~\cite{Bozek:2012gr,Qin:2013bha}, assuming strong interactions
between final state partons or hadrons, can give a satisfactory
description of the \pPb\ correlation data. However, the modeling of
small systems such as \pPb\ is complicated because uncertainties
related to initial state geometrical fluctuations play a large role
and because viscous corrections may be too large for hydrodynamics to
be a reliable framework~\cite{Bzdak:2013zma}. Additional experimental
information is therefore required to reveal the origin of these
correlations.  The \pt\ distributions and yields of particles of
different mass at low and intermediate momenta of less than a few
\gevc\ (where the vast majority of particles is produced), can provide
important information about the system created in high-energy hadron
reactions.

Previous results on identified particle production in pp~\cite{Aamodt:2011zj,Aamodt:2011zz, Abelev2012309,Chatrchyan:2012qb,Khachatryan:2011tm} and \PbPb~\cite{prl-spectra, Abelev:2013vea}
collisions at the LHC have been reported.  In this paper
we report on the measurement of \allpart\ production as a function
of the event multiplicity in \pPb\
collisions at a nucleon-nucleon center-of-mass energy
\snn~=~5.02~TeV. The results are presented over the following \pt\ ranges: 0.1-3, 0.2-2.5, 0-8,
0.3-4 and 0.6-8~\gevc\ for \allpart, respectively.  Results on $\pi$,
K, p production in \pPb\ collisions have been recently reported by the CMS collaboration~\cite{Chatrchyan:2013eya}.

\section{Sample and Data analysis}

\begin{table}[t] 
  \centering
  \begin{tabular*}{\linewidth}{@{\extracolsep{\fill}}ccc}
    \hline
    &&\\[-0.7em]
     Event & V0A range & $\avg{\dNdeta}$\\
     class & \footnotesize{(arb. unit)} & \footnotesize{$|\hlab|<0.5$}\\[0.3em]
    \hline
    &&\\[-0.7em]
    0--5\%    & $>$ 227  & 45   $\pm$ 1   \\[0.3em]
    5--10\%   & 187--227 & 36.2 $\pm$ 0.8 \\[0.3em]
    10--20\%  & 142--187 & 30.5 $\pm$ 0.7 \\[0.3em]
    20--40\%  & 89--142  & 23.2 $\pm$ 0.5 \\[0.3em] 
    40--60\%  & 52--89   & 16.1 $\pm$ 0.4 \\[0.3em]
    60--80\%  & 22--52   & 9.8  $\pm$ 0.2 \\[0.3em]
    80--100\% & $<$ 22   & 4.4  $\pm$ 0.1 \\[0.3em]
    \hline
  \end{tabular*}
  \caption{Definition of the event classes as fractions of the analyzed event sample and their corresponding $\avg{\dNdeta}$ within $|\hlab|<0.5$ (systematic uncertainties only, statistical uncertainties are negligible). }
  \label{tab:multclasses}
\end{table}

The results presented in this letter are obtained from a sample of the data
collected during the LHC \pPb\ run at \snn~=~5.02~TeV 
in the beginning of 2013. 
Because of the 2-in-1 magnet design of the LHC~\cite{Evans:2008zzb}, the energy of the two beams cannot be adjusted independently and is 4 ZTeV, leading to different energies due to the different Z/A. The nucleon-nucleon center-of-mass system, therefore, 
was moving in the laboratory frame with a rapidity of \ynn\ = $-0.465$ in the direction of the proton beam. 
The number of colliding bunches was varied
from 8 to 288. 
The total number of protons and Pb ions in the beams ranged from $0.2\times 10^{12}$ to $6.5\times 10^{12}$ and from $0.1\times 10^{12}$ to $4.4\times 10^{12}$, respectively. The maximum luminosity at the ALICE interaction point was for the data used in this paper 
$5\times 10^{27}\mathrm{cm}^{-2}\mathrm{s}^{-1}$ resulting in a
hadronic interaction rate of $10$ kHz.
The interaction region had an r.m.s. of 6.3~cm along the beam 
direction and of about 60~$\mu$m in the direction transverse to the beam.
For the results presented in this letter, a low-luminosity data sample has been analyzed where the event pile-up rate has been estimated to have negligible effects on the results. The integrated luminosity corresponding to the used data sample was about 14 $\rm \mu b^{-1}$ (7 $\rm \mu b^{-1}$) for the neutral (charged) hadron analysis. The LHC configuration was such that the lead beam circulated in the ``counter-clockwise'' direction, corresponding to the ALICE A direction or positive rapidity as per the convention used in this paper.

A detailed description of the ALICE apparatus can be found in \cite{Aamodt:2008zz}.
The minimum-bias trigger signal was provided by the VZERO counters, two arrays of 32 scintillator tiles each
covering the full azimuth within $2.8 < \hlab < 5.1$ (VZERO-A, Pb beam direction) and $-3.7 < \hlab < -1.7$ (VZERO-C, p beam direction). The signal amplitude and arrival time collected in each tile were recorded. A coincidence of signals in both VZERO-A and VZERO-C detectors was required to remove contamination from single diffractive and electromagnetic events~\cite{ALICE:2012xs}. The time resolution is better than 1 ns, allowing discrimination of beam--beam collisions from background events produced outside of the interaction region. In the offline analysis, background was further suppressed by the time information recorded in two neutron Zero Degree Calorimeters (ZDCs), which are located at $+112.5$ m (\ZNA) and $-112.5$ m (\ZNC) from the interaction point. 
A dedicated quartz radiator Cherenkov detector (T0) provided a measurement
of the event time of the collision.

\begin{table}[t]
  \centering
  \begin{tabular*}{\linewidth}{@{\extracolsep{\fill}}lr}
    \hline
    &\\[-0.7em]
    Selection variable & Cut value \\[0.3em]
    \hline
    &\\[-0.7em]
    2D decay radius & $>0.50$~cm \\[0.3em]
    Daughter track DCA to prim. vertex & $>0.06~$cm \\[0.3em]
    DCA between daughter tracks & $<1.0~\sigma$ \\[0.3em]
    Cosine of pointing angle (\kzero) & \pt\ dependent \\[0.3em]
    & ($<1\%$ signal loss) \\[0.3em]
    Cosine of pointing angle (\lmb\ and \almb) & \pt\ dependent \\[0.3em]
    & ($<1\%$ signal loss) \\[0.3em]
    Proper lifetime (\kzero) & $<20$~cm \\[0.3em]
    Proper lifetime (\lmb\ and \almb) & $<30$~cm \\[0.3em]   
    \kzero\ mass rejection window (\lmb\ and \almb) & $\pm 10$~\mevc\ \\[0.3em]
    \lmb\ and \almb\ mass rejection window (\kzero) & $\pm 5$~\mevc\ \\[0.3em]
    \hline
  \end{tabular*}
  \caption{\vzero\ topological selection cuts (DCA: distance-of-closest approach).}
  \label{tab:v0cuts}
\end{table}

The ALICE central-barrel tracking detectors cover the full azimuth
within $| \hlab |<0.9$.  They are located inside a solenoidal magnet
providing a magnetic field of 0.5 T. The innermost barrel detector is
the Inner Tracking System (ITS).  It consists of six layers of silicon
devices grouped in three individual detector systems which employ
different technologies (from the innermost outwards): the Silicon
Pixel Detector (SPD), the Silicon Drift Detector (SDD) and the Silicon
Strip Detector (SSD).  The Time Projection Chamber (TPC), the main
central-barrel tracking device, follows outwards.  Finally the
Transition Radiation Detector (TRD) extends the tracking farther away
from the beam axis. The primary vertex position was determined
separately in the SPD~\cite{ALICE:2012xs} and from tracks
reconstructed in the whole central barrel (global tracks).  The events were
further selected by requiring that the longitudinal position of the
primary vertex was within $10$~cm of the nominal interaction point and
that the vertices reconstructed from SPD tracklets and from global
tracks are compatible. In total from a sample of 29.8 (15.3) million
triggered events about 24.7 (12.5) million events passing the selection
criteria were used in the neutral (charged) hadron analysis.

\begin{table*}[t!]
  \centering 
  \begin{tabular*}{\linewidth}{@{\extracolsep{\fill}}lcccccc}
    \hline
    &&&&&&\\[-0.7em]
     & \multicolumn{2}{c}{$\pi^{\pm}$} & \multicolumn{2}{c}{K$^{\pm}$} & \multicolumn{2}{c}{p(\pbar)} \\[0.3em]
    \hline
    \hline
    &&&&&&\\[-0.7em]
    \pt\ (\gevc) & 0.1 & 3 & 0.2 & 2.5 & 0.3 & 4\\[0.3em]
    \hline
    &&&&&&\\[-0.7em]
    Correction for & \multirow{2}{*}{1\%} & \multirow{2}{*}{1\%} & \multicolumn{2}{c}{\multirow{2}{*}{negl.}} & \multirow{2}{*}{4\%} & \multirow{2}{*}{1\%} \\ 
    secondaries & & & & & & \\[0.3em]
    
    Material & \multirow{2}{*}{5\%} & \multirow{2}{*}{negl.} & \multirow{2}{*}{2.5\%} & \multirow{2}{*}{negl.} & \multirow{2}{*}{4\%} & \multirow{2}{*}{negl.} \\
    budget & & & & & & \\[0.3em]

    Hadronic      &  \multirow{2}{*}{2\%} & \multirow{2}{*}{1\%} &   \multirow{2}{*}{3\%} & \multirow{2}{*}{1\%} &  
    6\% & 1\% (\pbar) \\
    interactions & & & & &  4\% & negl. (p) \\[0.3em]
    
    Global tracking&  \multicolumn{2}{c}{\multirow{2}{*}{4\%}} & \multicolumn{2}{c}{\multirow{2}{*}{4\%}} & \multicolumn{2}{c}{\multirow{2}{*}{4\%}} \\
    efficiency &&&&&& \\[0.3em]
    
    Multiplicity&  \multirow{2}{*}{2\%} & \multirow{2}{*}{negl.} & \multirow{2}{*}{4\%} &\multirow{2}{*}{negl.} & \multirow{2}{*}{2\%}  & \multirow{2}{*}{negl.} \\
    dependence &&&&&& \\[0.3em]

    \hline
    \hline
    &&&&&&\\[-0.7em]
    \pt\ (\gevc) & 0.1 & 0.6 & 0.2 & 0.5 & 0.3 & 0.6\\[0.3em]
    \hline
    &&&&&&\\[-0.7em]
    ITS standalone&  \multirow{2}{*}{5\%} & \multirow{2}{*}{4\%} & \multirow{2}{*}{6\%} & \multirow{2}{*}{4.5\%} & \multirow{2}{*}{6\%} & \multirow{2}{*}{4.5\%} \\ 
    tracking efficiency &&&&&& \\[0.3em]
    
    ITS PID & \multicolumn{2}{c}{1\%} & \multicolumn{2}{c}{2\%} & \multicolumn{2}{c}{1.5\%} \\[0.3em]

    \hline
    \hline
    &&&&&&\\[-0.7em]
    \pt\ (\gevc) & 0.3 & 0.65 & 0.3 & 0.6 & 0.5 & 0.9\\[0.3em]
    \hline
    &&&&&&\\[-0.7em]
    
    TPC PID & \multicolumn{2}{c}{1.5\%} & \multicolumn{2}{c}{3.5\%} & \multicolumn{2}{c}{2.5\%} \\[0.3em]
    
    \hline
    \hline
    &&&&&&\\[-0.7em]
    \pt\ (\gevc) & 0.5 & 3 & 0.5 & 2.5 & 0.5 & 4\\[0.3em]
    \hline
    &&&&&&\\[-0.7em]
    
    TOF matching &  \multirow{2}{*}{4\%} & \multirow{2}{*}{3\%} & \multirow{2}{*}{5\%} &\multirow{2}{*}{4\%} & \multirow{2}{*}{5\%}  & \multirow{2}{*}{3\%} \\
    efficiency &&&&&& \\[0.3em]

    TOF PID & 1\% & 10\% & 2\% & 17\% & 2\% & 20\% \\[0.3em]
    \hline
    \hline
    &&&&&&\\[-0.7em]
    \pt\ (\gevc) & 0.1 & 3 & 0.2 & 2.5 & 0.3 & 4\\[0.3em]
    \hline
    &&&&&&\\[-0.7em]
    Total & 7.5\% & 12\% & 8.5\% & 20\% & 9.5\% & 20\% \\[0.3em]
    \hline
    
  \end{tabular*}
  
  \caption{Main sources of systematic uncertainty for \allpi, \allk, \allp.} \label{tab:syst}

\end{table*}

In order to study the multiplicity dependence, the selected event
sample was divided into seven event classes, based on cuts on the
total charge deposited in the \VZEROA\ detector (V0A).
The corresponding fractions of the data sample in each class
are summarized in Tab.~\ref{tab:multclasses}.  
The mean charged-particle multiplicity densities~($\avg{\dNdeta}$)
within $| \hlab |<0.5$ corresponding to the different centrality bins
are also listed in the table.  These are obtained using the method
presented in~\cite{ALICE:2012xs} and are corrected for acceptance and
tracking efficiency as well as for contamination by secondary
particles. The relative standard deviation of the track multiplicity
distribution for the event classes defined in
Table~\ref{tab:multclasses} ranges from 78\% to 29\% for
the 80--100\% and 0--5\% classes, respectively. It should be noted
that the average multiplicity in the 80-100\% bin is well below the
corresponding multiplicity in pp minimum-bias
collisions~\cite{Aamodt:1260702} and therefore likely to be subject to
a strong selection bias.  Contrary to our
earlier measurement of $\avg{\dNdeta}$~\cite{ALICE:2012xs}, the values
in Tab.~\ref{tab:multclasses} are not corrected for trigger and
vertex-reconstruction efficiency, which is of the order of 2\% for NSD
events~\cite{ALICE:2012xs}. The same holds true for the \pt\
distributions, which are presented in the next section.

Charged-hadron identification in the central barrel was performed with
the ITS, TPC~\cite{Alme:2010ke} and Time-Of-Flight
(TOF)~\cite{Akindinov:2013tea} detectors. The drift and strip layers
of the ITS provide a measurement of the specific energy loss with a
resolution of about 10\%. In a standalone tracking mode, the
identification of pions, kaons, and protons is thus extended down to
respectively 0.1, 0.2, 0.3 \gevc\ in \pt.  The TPC provides particle
identification at low momenta via specific energy loss \dedx\ in the
fill gas by measuring up to 159 samples per track with a resolution of
about 6\%. The separation power achieved in \pPb\ collisions is
identical to that in pp
collisions~\cite{performance-paper}. Further outwards at about 3.7 m
from the beam line, the TOF array allows identification at higher \pt\
measuring the particle speed with the time-of-flight technique. The
total time resolution is about 85 ps for events in the multiplicity
classes from 0\% to $\sim 80$\%.  In more peripheral collisions, where
multiplicities are similar to pp, it decreases to about 120 ps due to
a worse start-time (collision-time)
resolution~\cite{performance-paper}.  The start-time of the event was
determined by combining the time estimated using the particle arrival
times at the TOF and the time measured by the T0
detector~\cite{Akindinov:2013tea}.

Since the \pPb\ center-of-mass system moved in the laboratory frame
with a rapidity of \ynn\ = $-0.465$, the nominal acceptance of the
central barrel of the ALICE detector was asymmetric with respect to
\ycms\ = 0.  In order to ensure good detector acceptance and optimal
particle identification performance, tracks were selected in the
rapidity interval $0 < \ycms < 0.5$ in the nucleon-nucleon
center-of-mass system. Event generator studies and repeating the
analysis in $\left|\ycms\right| < 0.2$ indicate differences between
the two rapidity selections smaller than 2\% in the
normalization and 3\% in the shape of the transverse momentum
distributions.

In this paper we present results for primary particles, defined as all particles produced in the collision, including decay products, but excluding weak decays of strange particles. The analysis technique is described in detail
in~\cite{prl-spectra, Abelev:2013vea, ALICE:2013xaa}. Here we briefly review
the most relevant points.  

\begin{table}[t]
\centering 
\begin{tabular*}{\linewidth}{@{\extracolsep{\fill}}lccc}
\hline
&&&\\[-0.7em]
 & \kzero\ & \multicolumn{2}{c}{\lmb(\almb)}\\[0.3em]
\hline
&&&\\[-0.7em]
Proper lifetime & 2\% & \multicolumn{2}{c}{2\%} \\[0.3em]
Material budget & 4\% & \multicolumn{2}{c}{4\%} \\[0.3em]
Track selection  & 4\% & \multicolumn{2}{c}{4\%} \\[0.3em]
TPC PID & 1\% & \multicolumn{2}{c}{1\%} \\[0.3em]
Multiplicity & \multirow{2}{*}{2\%} & \multicolumn{2}{c}{\multirow{2}{*}{2\%}} \\
dependence & & \\[0.3em]
\hline
\hline
&&&\\[-0.7em]
\pt\ (\gevc)  &  & $<$ 3.7 & $>$ 3.7\\[0.3em]
\hline
&&&\\[-0.7em]
Feed-down  &  & \multirow{2}{*}{5\%} & \multirow{2}{*}{7\%}\\
correction & & &\\[0.3em]
    \hline
    \hline
    &&&\\[-0.7em]
\pt\ (\gevc)  &  & $<$ 3.7 & $>$ 3.7\\[0.3em]
    \hline
    &&&\\[-0.7em]
    Total & 6.5\% & 8\% & 9.5\% \\[0.3em]
\hline
\end{tabular*}
\caption{Main sources of systematic uncertainty for the \kzero\ and \lmb(\almb).} \label{tab:v0syst}
\end{table}

Three approaches were used for the identification of $\pi^{\pm}$, K$^{\pm}$, and p($\bar{\rm p}$), called ``ITS standalone'', ``TPC/TOF'' and ``TOF
fits''~\cite{prl-spectra, Abelev:2013vea} in the following. In the ``ITS standalone'' method, a probability for each particle species is calculated in each layer based on the measured energy loss signal and the known response function. The information from all layers is combined in a bayesian approach with iteratively determined priors. Finally, the type with the highest probability is assigned to the track. This method is used in the \pt\ ranges 
0.1 $<$ \pt\ $<$ 0.7 \gevc, 0.2 $<$ \pt\ $<$ 0.6 \gevc\ and 0.3 $<$ \pt\ $<$ 0.65 \gevc\ for $\pi^{\pm}$, K$^{\pm}$, and p($\bar{\rm p}$), respectively. In contrast to the analysis in the high multiplicity environment of central heavy-ion collisions, the contribution of tracks with wrongly associated clusters is negligible in \pPb\ collisions. In the ``TPC/TOF'' method, the
particle is identified by requiring that its measured \dedx\ and time-of-flight
are within $\pm$3$\sigma$ from the expected values in the TPC and/or TOF.
This method is used in the \pt\ ranges 
0.2 $<$ \pt\ $<$ 1.5 \gevc, 0.3 $<$ \pt\ $<$ 1.3 \gevc\ and 
0.5 $<$ \pt\ $<$ 2.0 \gevc\ for $\pi^{\pm}$, K$^{\pm}$, and p($\bar{\rm p}$),
respectively.  In the third method the TOF time distribution is fitted
to extract the yields, with the expected shapes based on the knowledge
of the TOF response function for different particle species. This
method is used in the \pt\ range starting from 0.5 \gevc\ up to 3, 2.5
and 4 \gevc\ for $\pi^{\pm}$, K$^{\pm}$, and p($\bar{\rm p}$), respectively.  Contamination from
secondary particles was subtracted with a data-driven approach, based
on the fit of the transverse distance-of-closest approach to the
primary vertex (DCA$_{xy}$)
distribution with the expected shapes for primary and secondary
particles~\cite{prl-spectra, Abelev:2013vea}. The results of the three analyses were combined using the (largely independent) systematic uncertainties as weights in the overlapping ranges, after checking for their compatibility.

The \kzero\ and \lmb(\almb) particles were identified exploiting their
``\vzero'' weak decay topology in the channels $\kzero \to \pi^{+}
\pi^{-}$ and $\lmb(\almb) \to \rm{p} \pi^{-} (\rm{\bar{p}} \pi^{+})$,
which have branching ratios of 69.2\% and 63.9\%,
respectively~\cite{Beringer:1900zz}. The selection criteria used to
define two tracks as \vzero\ decay candidates are listed in
Tab.~\ref{tab:v0cuts} (see~\cite{Aamodt:2011zz} for details). Since
the cosine of pointing angle (the angle between the particle momentum associated with the V0 candidate and a vector connecting the primary vertex and the V0 position~\cite{Aamodt:2011zz}) resolution changes significantly with
momentum, the value used in the selection is \pt\ dependent and such
that no more than 1\% of the primary particle signal is removed.

 The typical reconstruction
efficiencies (excluding branching ratios) are about 15\% at low \pt\
($\sim$ 0.5 \gevc), increasing to about 70\% for \kzero\ and 55\% for
\lmb(\almb) at higher momenta (\pt\ $>$ 3 \gevc).  The signal is
extracted from the reconstructed invariant mass distribution
subtracting the background from the peak region with a bin counting
method.  The background and signal regions are defined on the basis of the
mass resolution as the windows in $\left[ -12 \sigma, -6 \sigma
\right]$, $\left[ 6 \sigma, 12 \sigma \right]$ and $\left[ -6 \sigma,
  6 \sigma \right]$, respectively. The value of $\sigma$ changes with
\pt\ to account for the actual mass resolution and ranges from about 3 MeV/$c^{2}$ to 7 MeV/$c^{2}$ for \kzero\
and from about 1.4 MeV/$c^{2}$ to 2.5 MeV/$c^{2}$ for \lmb(\almb).
More details on \vzero\ reconstruction can be found
in~\cite{Aamodt:2011zz, ALICE:2013xaa}.  The contribution from weak decays
of the charged and neutral $\Xi$ to the \lmb(\almb) yield has been
corrected following a data-driven approach.  The measured $\Xi^{-} (\bar{\Xi}^+)$
spectrum is used as input in a simulation of the decay kinematics to
evaluate the fraction of reconstructed \lmb(\almb) coming from
$\Xi^{-} (\bar{\Xi}^+)$ decays.  The contribution from the decays of $\Xi^{0}$ is
taken into account in the same way by assuming the ratio
$\Xi^{-} (\bar{\Xi}^+)/\Xi^{0} = 1$, as supported by statistical models and Pythia
or DMPJET Monte Carlo simulations~\cite{Roesler:2000he,
  Skands:2010ak}.
The raw transverse momentum distributions have been corrected for acceptance and
reconstruction efficiency using a
Monte Carlo simulation, based on the DPMJET 3.05 event
generator~\cite{Roesler:2000he} and a GEANT3.21~\cite{Brun:1994aa} model of the
detector. As compared to the version used in~\cite{prl-spectra, Abelev:2013vea},
GEANT3.21 was improved by implementing a more realistic parameterization of the
anti-proton inelastic cross-section~\cite{eulogio-papers}.
A correction factor based on FLUKA~\cite{Battistoni:2007zzb} estimates 
was applied to negative kaons as in \cite{prl-spectra, Abelev:2013vea}.

The study of systematic uncertainties follows the analysis described
in~\cite{prl-spectra, Abelev:2013vea} for $\pi^{\pm}$, K$^{\pm}$ and p($\bar{\rm p}$). The main
sources are the correction for secondary particles (4\% for protons, 1\% for
pions, negligible for kaons), knowledge of the material budget 
(3\% related to energy
loss), hadronic interactions with the detector material (from 1\% to 6\%,
more important at low \pt\ and for protons), tracking efficiency (4\%),
TOF matching efficiency (from 3 to 6\%, depending on the particle) and PID
(from 2\% to 25\%, depending on the particle and the \pt\ range).  For
the neutral \lmb\ and \kzero\ particles, the main sources are the level of 
knowledge of detector
materials (resulting in a 4\% uncertainty), track selections (up to 5\%) and the
feed-down correction for the $\Lambda$ and $\bar{\Lambda}$ (5\%), while
topological selections contribute 2-4\% depending on transverse
momentum. The main sources of systematic uncertainties for the
analysis of charged and neutral particles are summarized in
Tables~\ref{tab:syst} and~\ref{tab:v0syst}, respectively.  The study of
systematic uncertainties was repeated for the different multiplicity
bins in order to separate the sources of uncertainty which are 
dependent on multiplicity and uncorrelated across different
bins (depicted as shaded boxes in the figures).

\section{Results}

\ifplb
\begin{figure*}[p]
\else
\begin{figure*}[t!]
\fi
  \begin{flushleft}
    \includegraphics[width=0.495\textwidth]{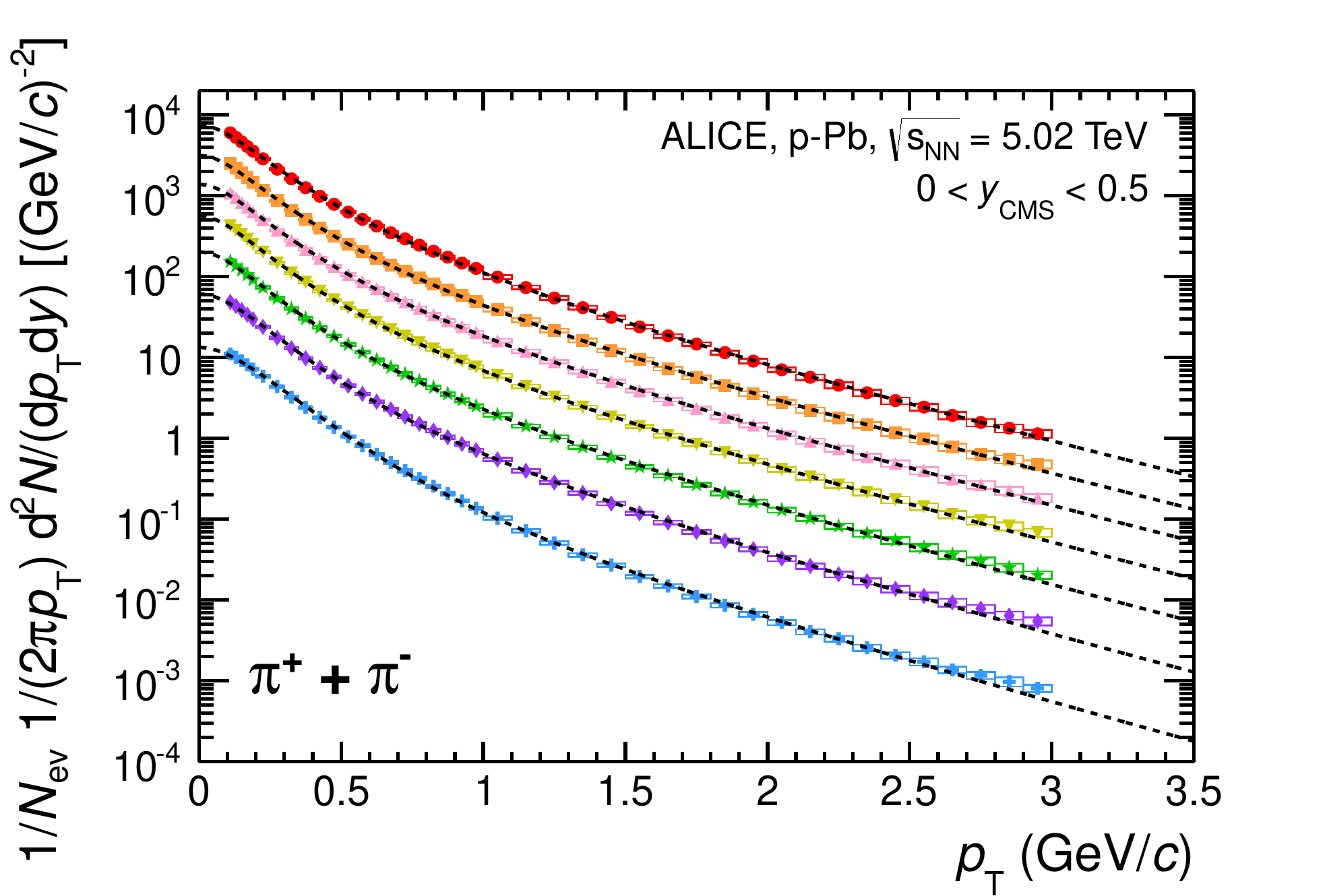}
    \includegraphics[width=0.495\textwidth]{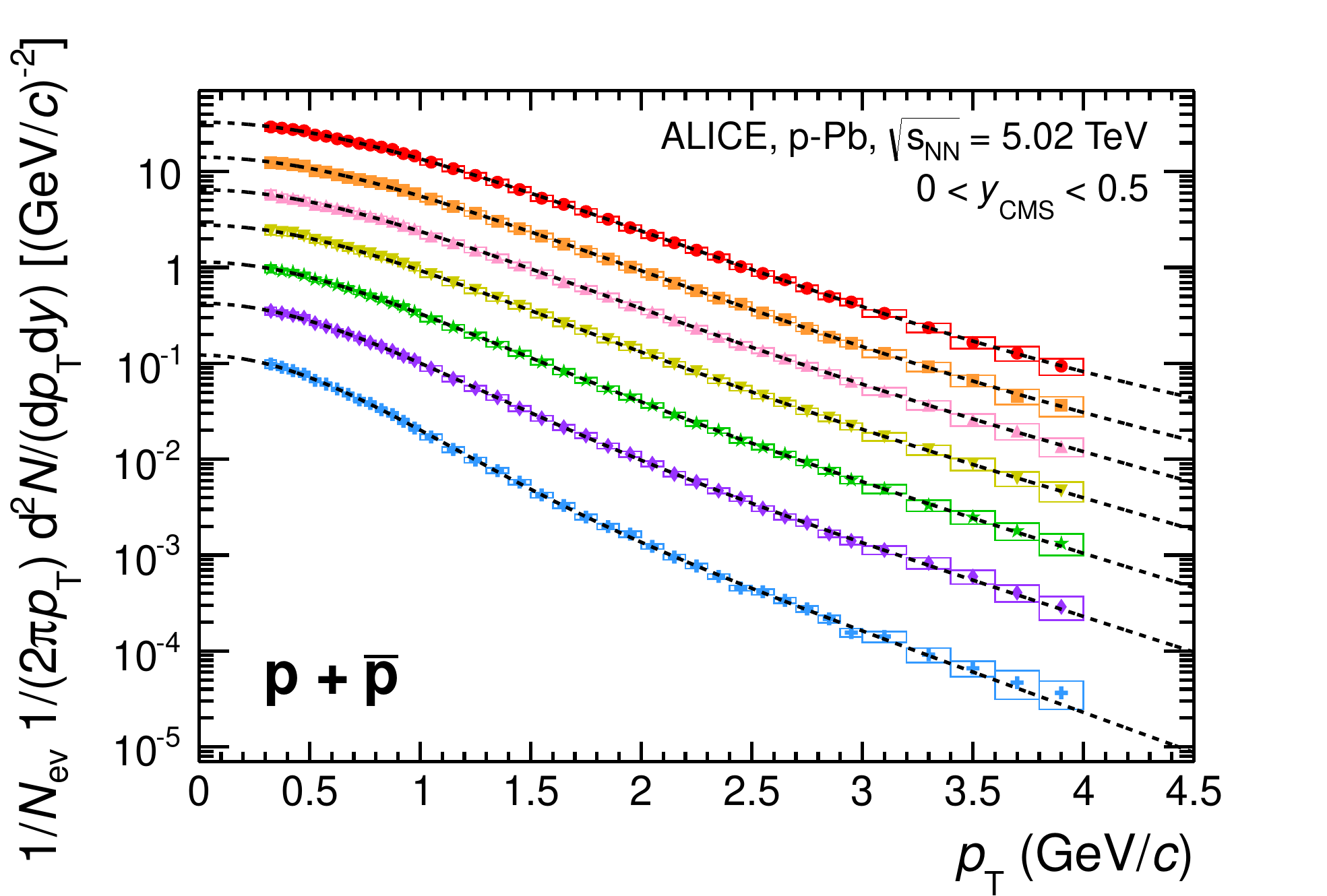}
    \includegraphics[width=0.495\textwidth]{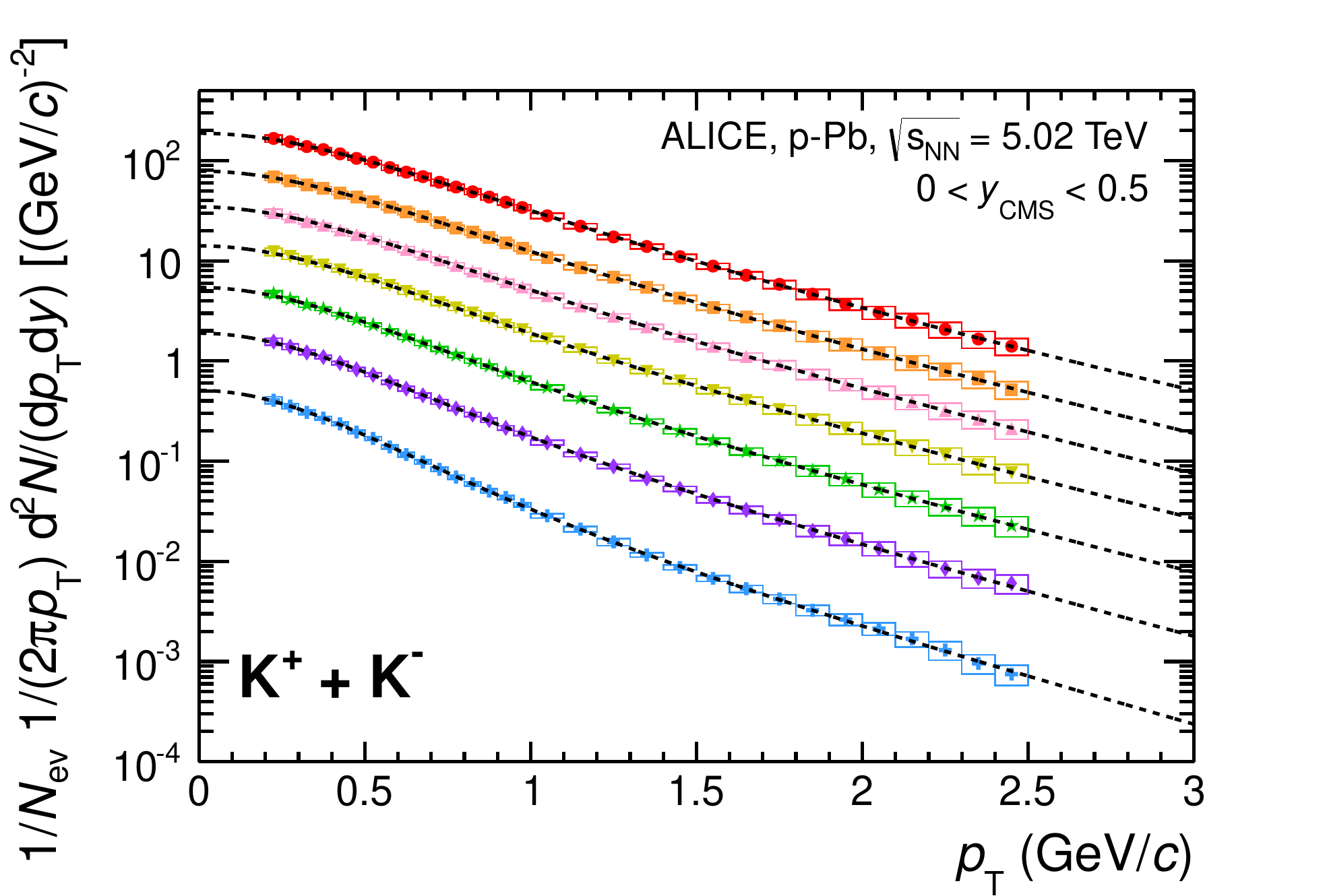}
    \includegraphics[width=0.495\textwidth]{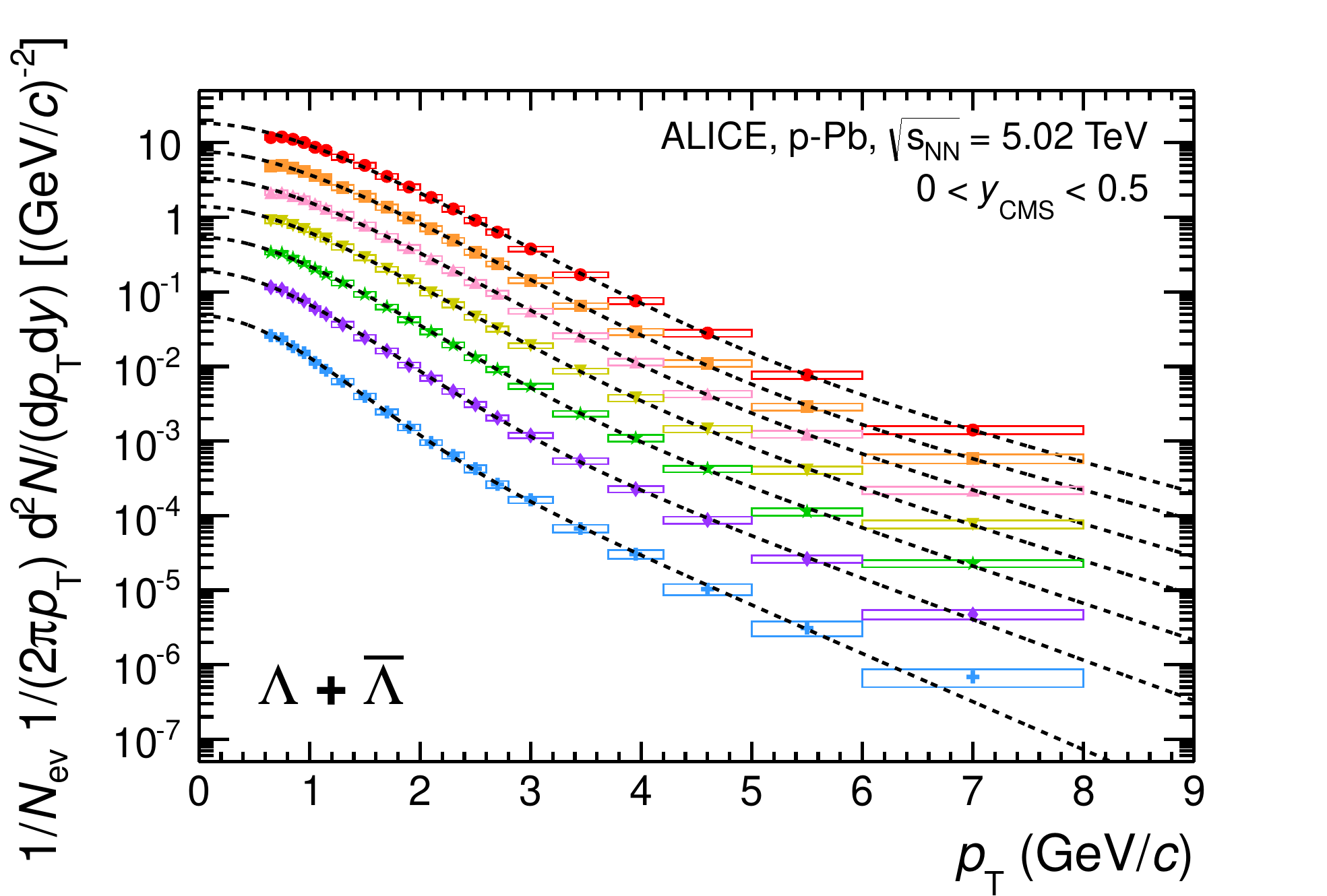}
    \includegraphics[width=0.495\textwidth]{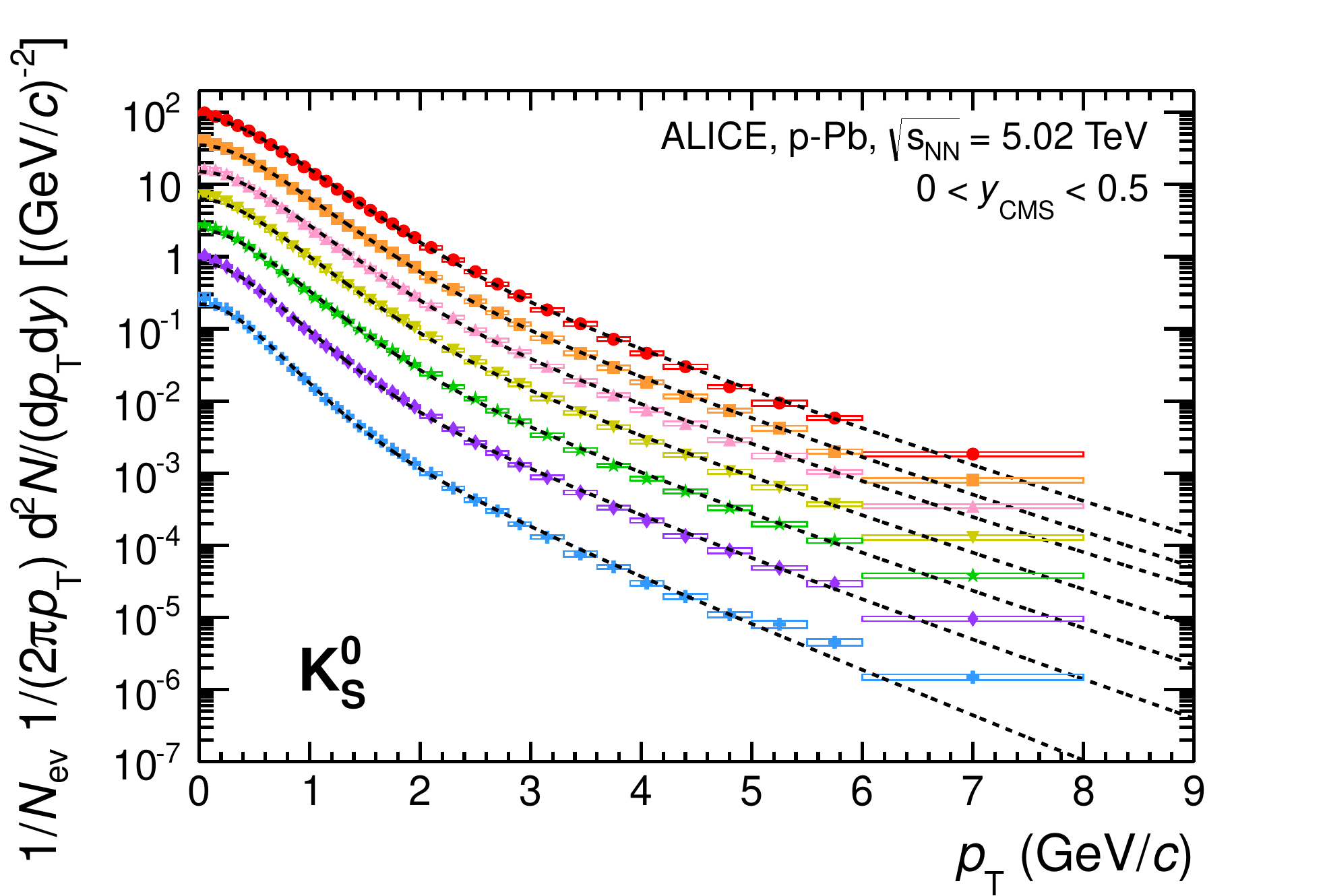}
    \includegraphics[width=0.495\textwidth]{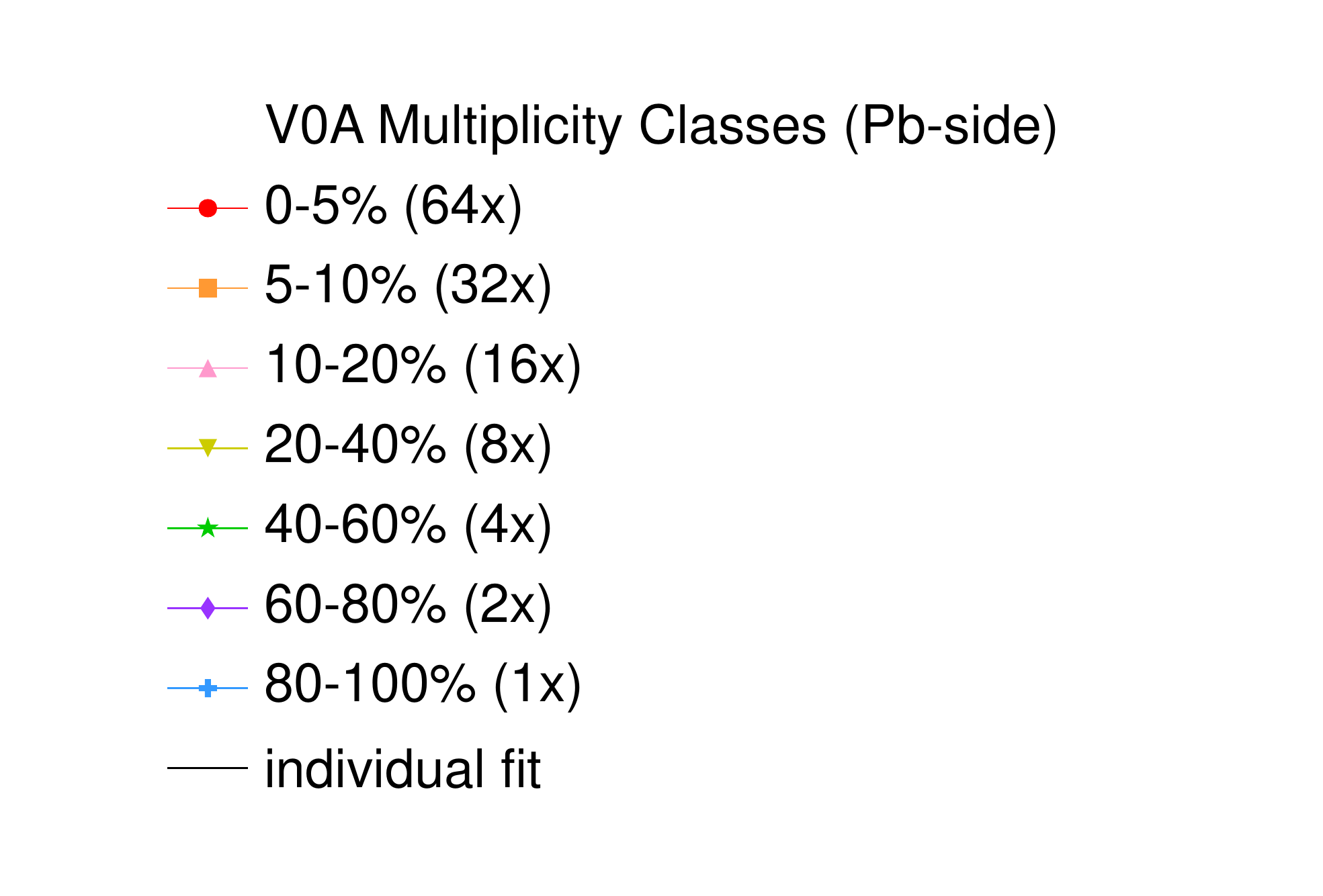}
    \end{flushleft}
    \caption{(color online) Invariant \pt-differential yields of \allpart\ in different V0A multiplicity classes (sum of particle and antiparticle states where relevant) measured in the rapidity interval $0 < \ycms < 0.5$. Top to bottom: central to peripheral; data scaled by $2^{n}$ factors for better visibility. Statistical (bars) and full systematic (boxes) uncertainties are plotted. Dashed curves: blast-wave fits to each individual distribution. }
    \label{fig:spectra}
\end{figure*}  

\ifplb
\begin{figure}[p]
  \centering
  \includegraphics[width=0.55\textwidth]{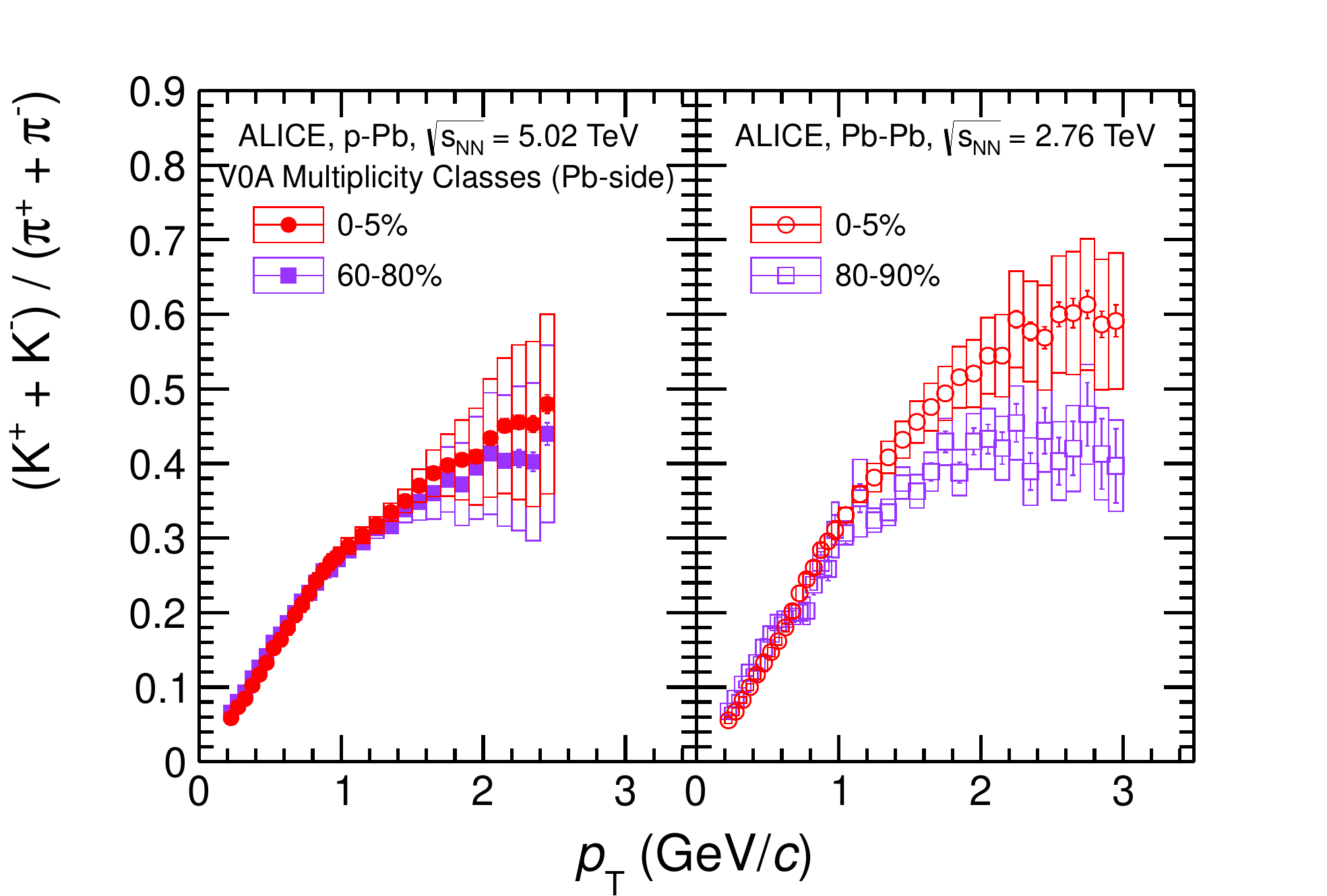}
  \includegraphics[width=0.55\textwidth]{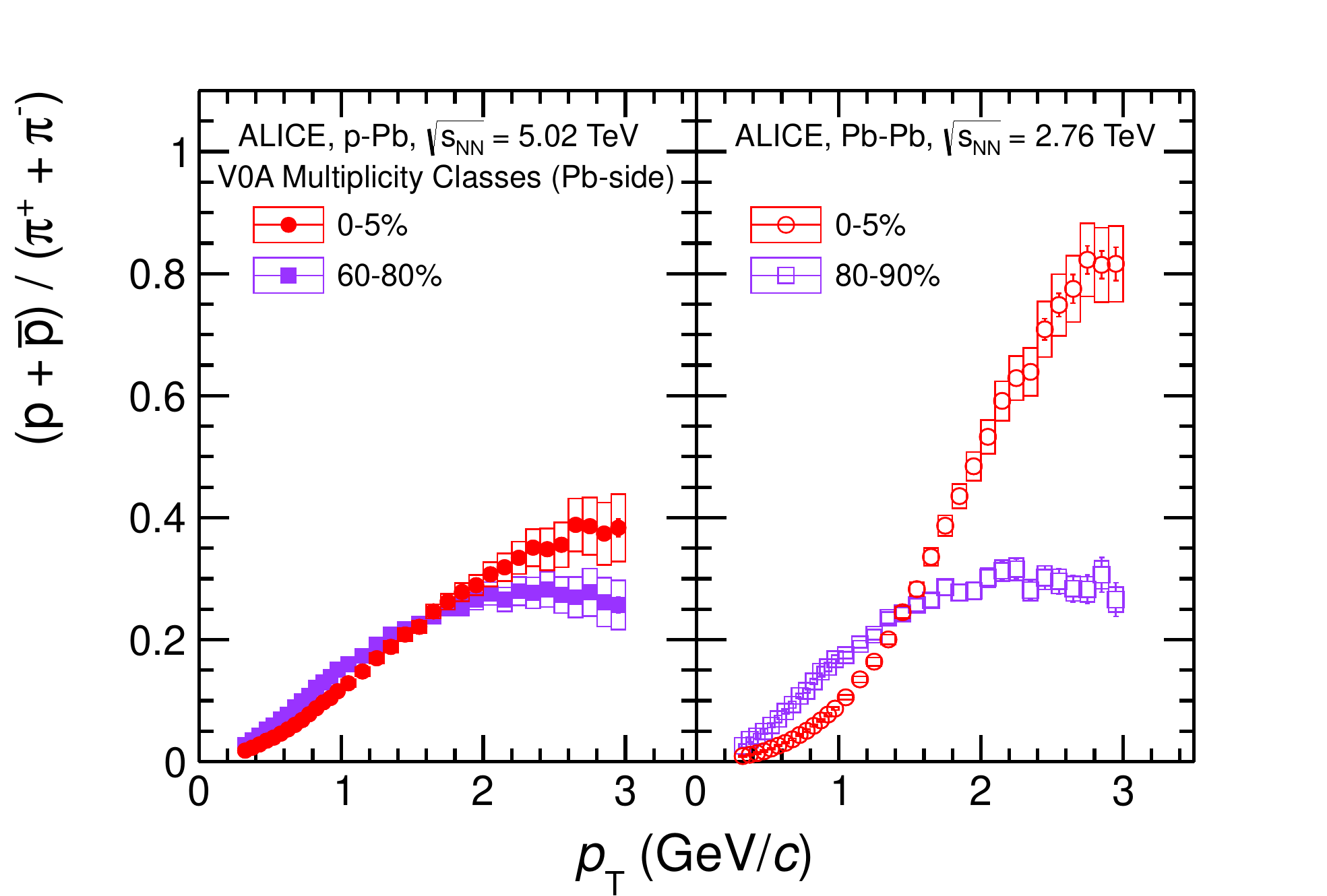}
  \includegraphics[width=0.55\textwidth]{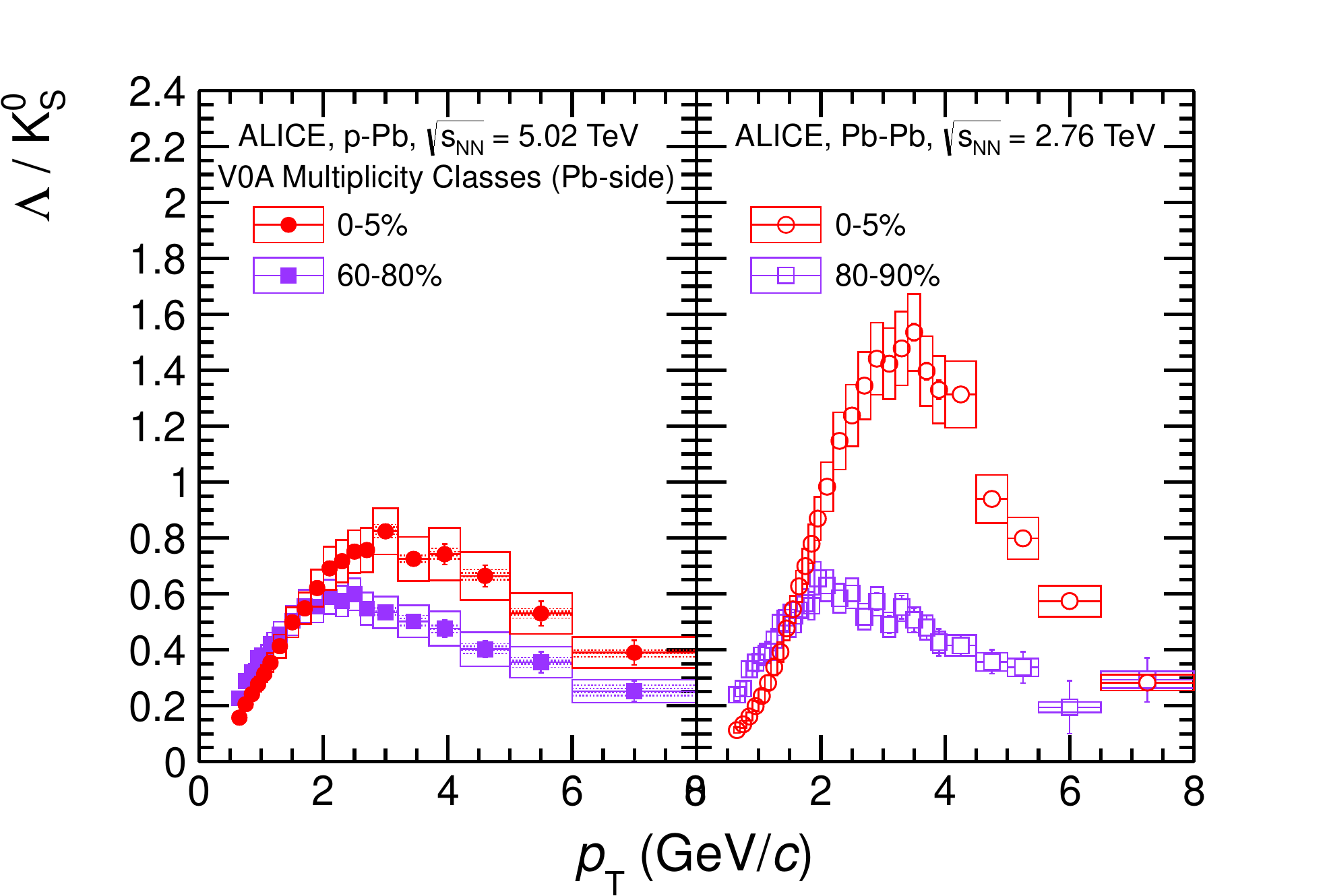}
\else
\begin{figure}[t!]
  \centering
  \includegraphics[width=0.5\textwidth]{KaonPionRatioPt}
  \includegraphics[width=0.5\textwidth]{ProtonPionRatioPt}
  \includegraphics[width=0.5\textwidth]{LambdaK0sRatioPt}
\fi
  \caption{(color online) Ratios \kpi~=~(\kap + \kam)/(\pip + \pim),  \ppi~=~(p + \pbar)/(\pip +
\pim) and  \lmb/\kzero\ as a function of \pt\ in two multiplicity bins measured in the rapidity interval $0 < \ycms < 0.5$  (left panels). The ratios are compared to results in Pb--Pb collisions measured at midrapidity, shown in the right panels. The empty boxes show the total systematic uncertainty; the shaded boxes indicate the contribution uncorrelated across multiplicity bins  (not estimated in Pb--Pb).}
  \label{fig:RatiosVsPt}
\end{figure}

\ifplb
\begin{figure}[p]
  \centering
  \includegraphics[width=0.55\textwidth]{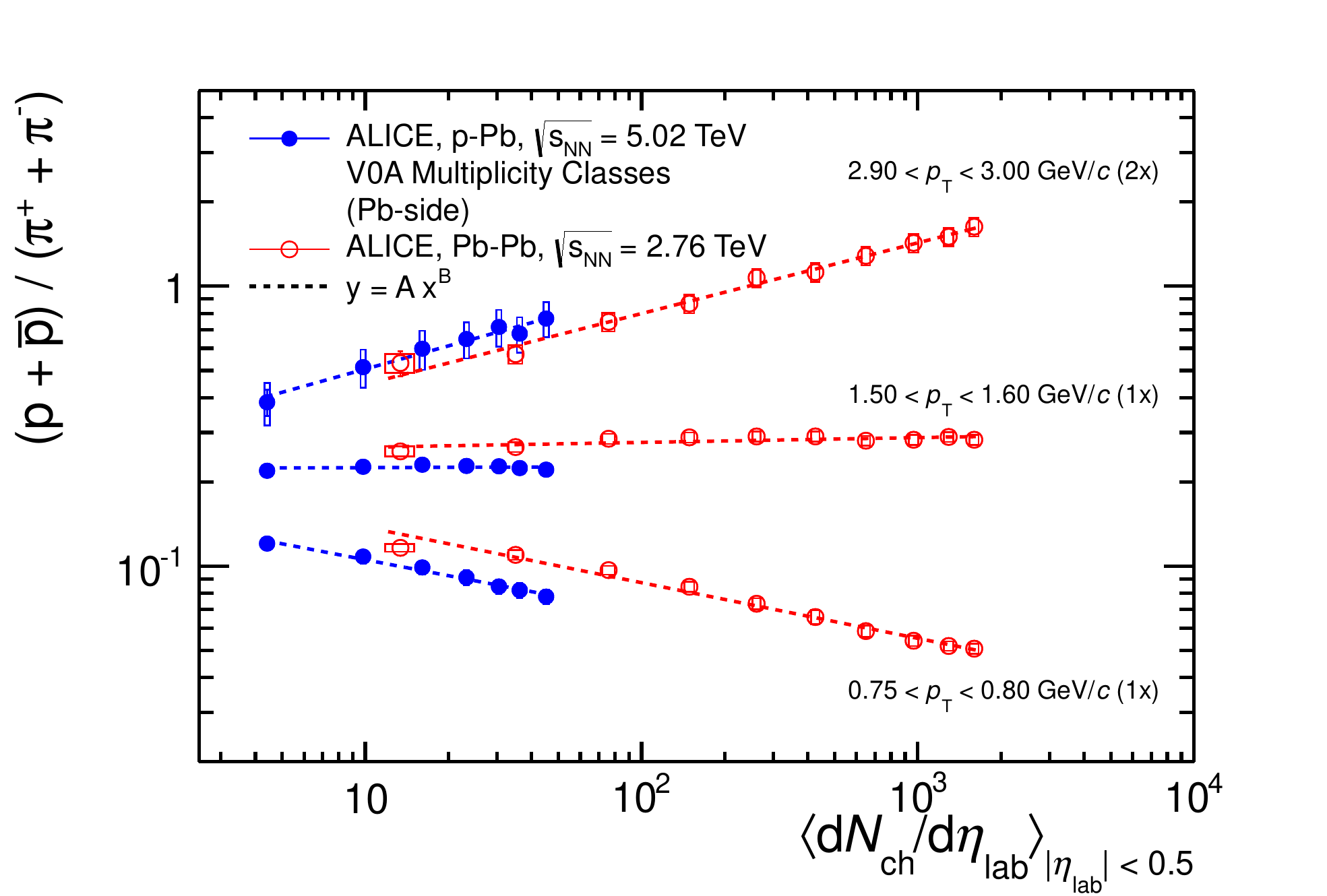}
  \includegraphics[width=0.55\textwidth]{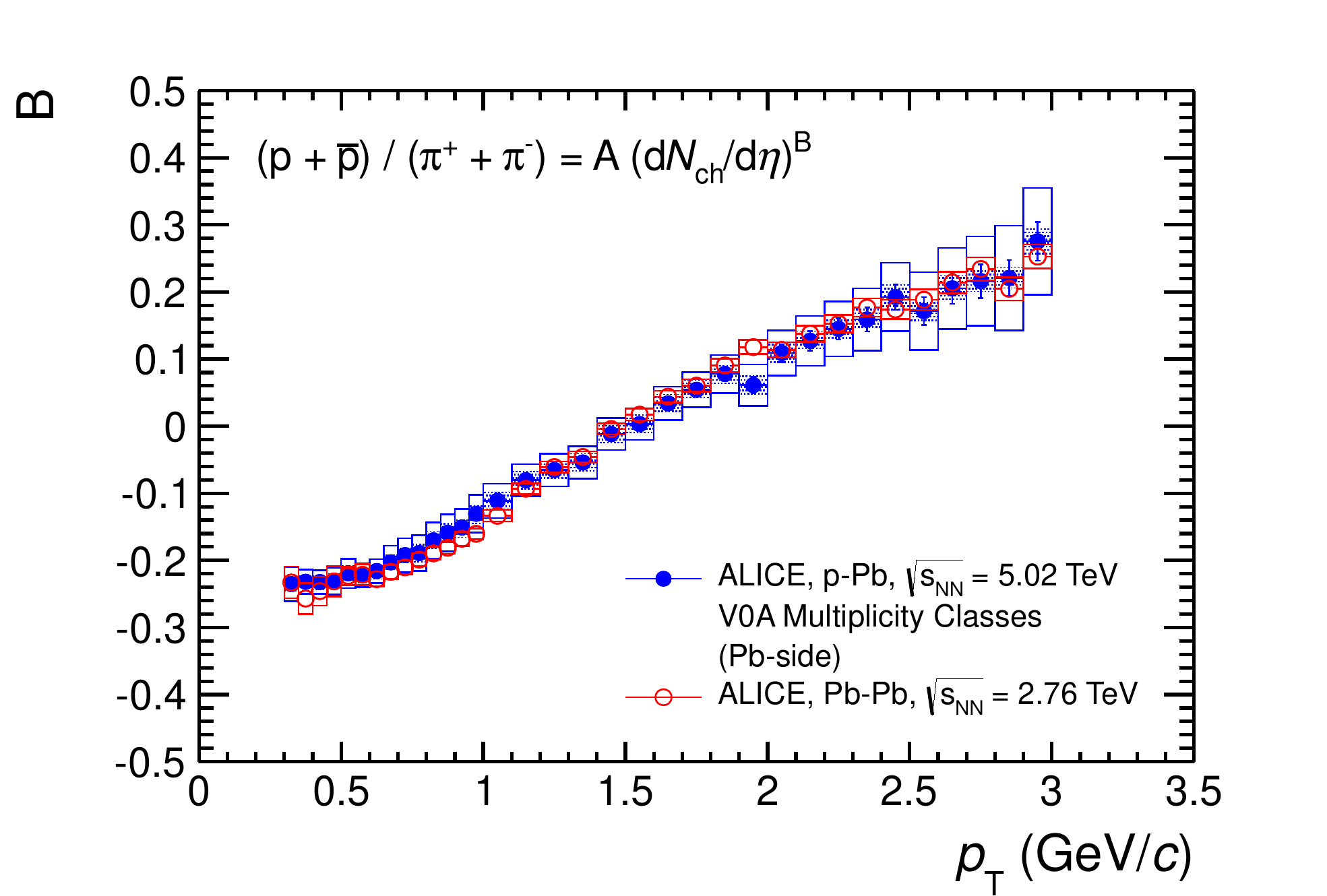}
  \includegraphics[width=0.55\textwidth]{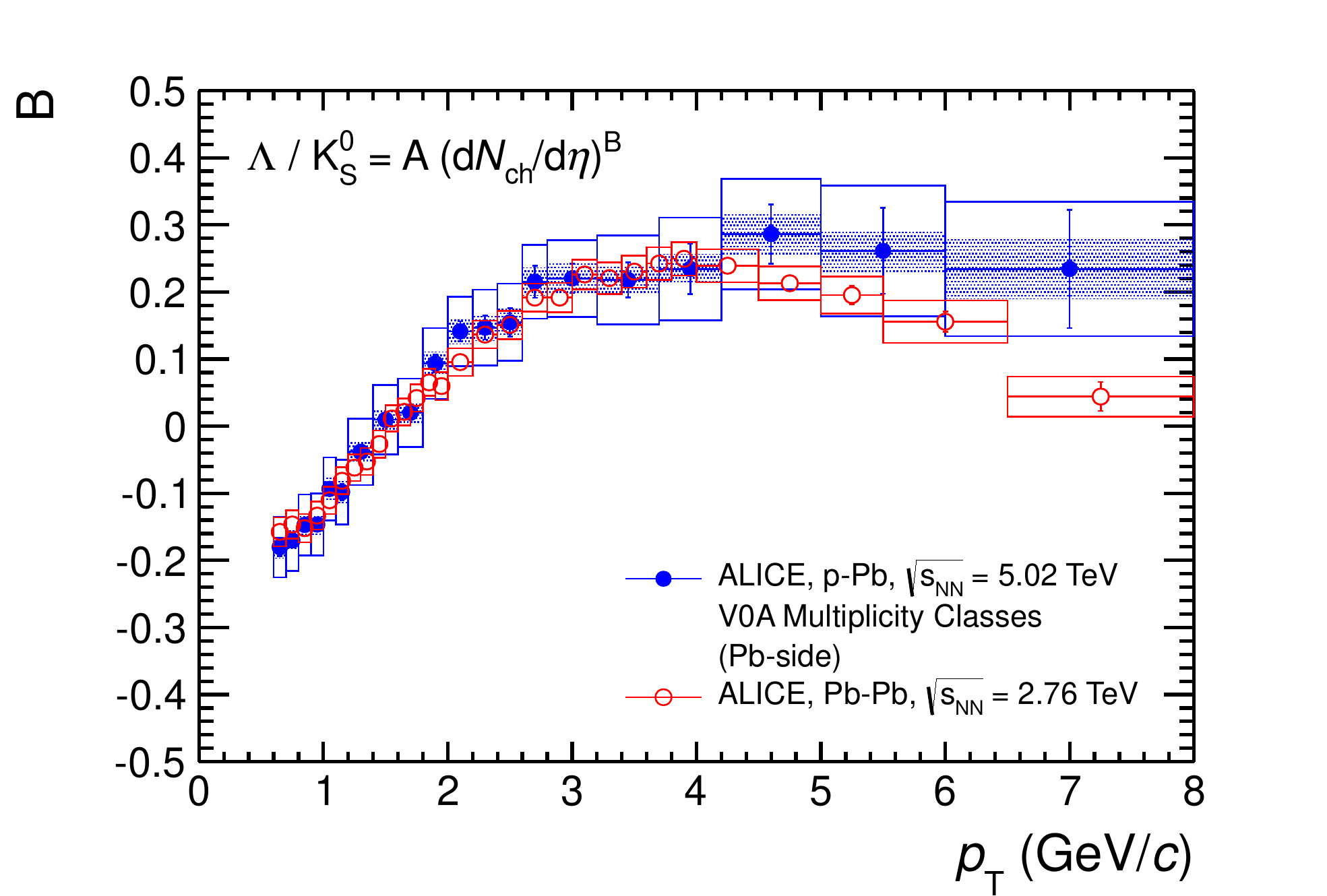}
\else
\begin{figure}[t!]
  \centering
  \includegraphics[width=0.5\textwidth]{ProtonPionRatioScaling}
  \includegraphics[width=0.5\textwidth]{ProtonPionRatioPower}
  \includegraphics[width=0.5\textwidth]{LambdaK0sRatioPower}
\fi 
 \caption{(color online) \ppi\ ratio as a function of the 
    charged-particle density
    \dNdeta\ in three \pt\ intervals in \pPb\ (measured in the rapidity interval $0 < \ycms < 0.5$) and \PbPb\ collisions (measured at midrapidity). The dashed lines show the corresponding power-law fit (top). Exponent of the \ppi\ (middle) and \lmb/\kzero\ (bottom) power-law fit as a function of \pt\ in \pPb\ and \PbPb\ collisions. The empty boxes show the total systematic uncertainty; the shaded boxes indicate the contribution uncorrelated across multiplicity bins  (not estimated in Pb--Pb).}
    \label{fig:p-scaling}
\end{figure}

\ifplb
\begin{figure}[t!]
  \centering
  \includegraphics[width=0.7\textwidth]{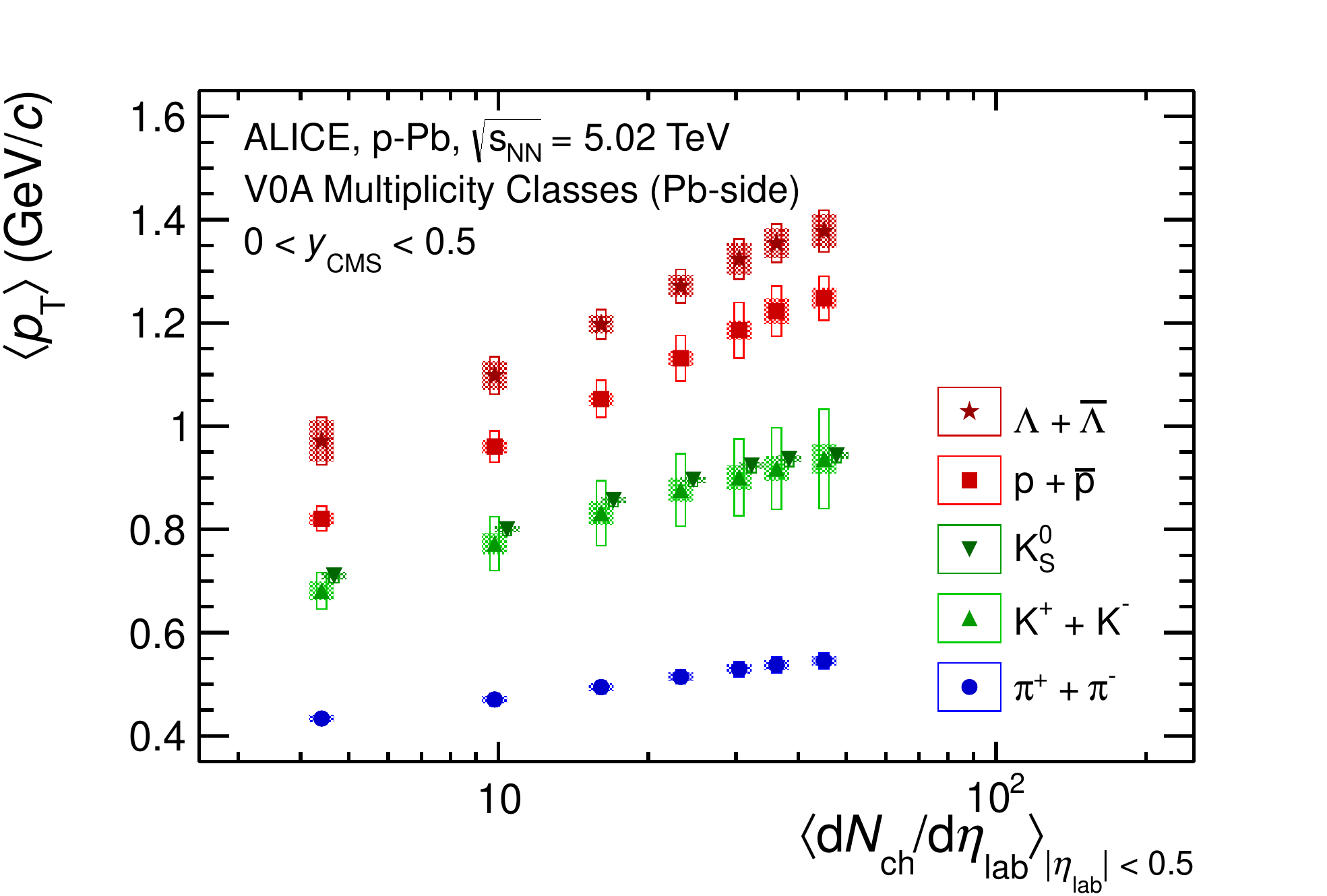}  
\else
\begin{figure}[t!]
  \centering
  \includegraphics[width=0.5\textwidth]{MeanPtAllParticles}  
\fi
  \caption{(color online) Mean transverse momentum as a function of \dNdeta\ in each V0A multiplicity class (see text for details) for different particle species measured in the rapidity interval $0 < \ycms < 0.5$. The \dNdeta\ values of \kzero\ are shifted for visibility. The empty boxes show the total systematic uncertainty; the shaded boxes indicate the contribution uncorrelated across multiplicity bins  (not estimated in Pb--Pb).}
  \label{fig:meanpt}
\end{figure}

\ifplb
\begin{figure}[p]
  \centering
  \includegraphics[width=0.55\textwidth]{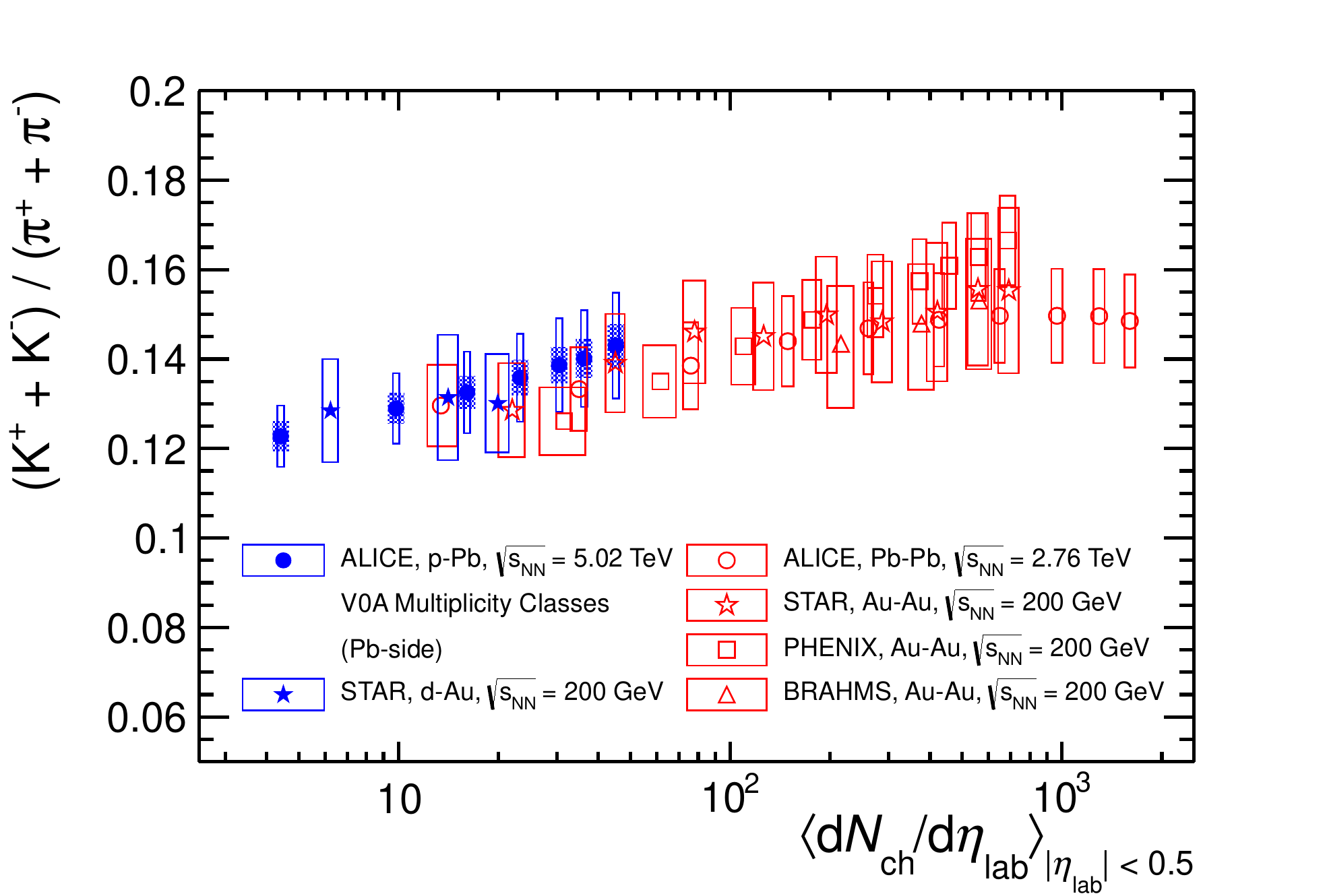}  
  \includegraphics[width=0.55\textwidth]{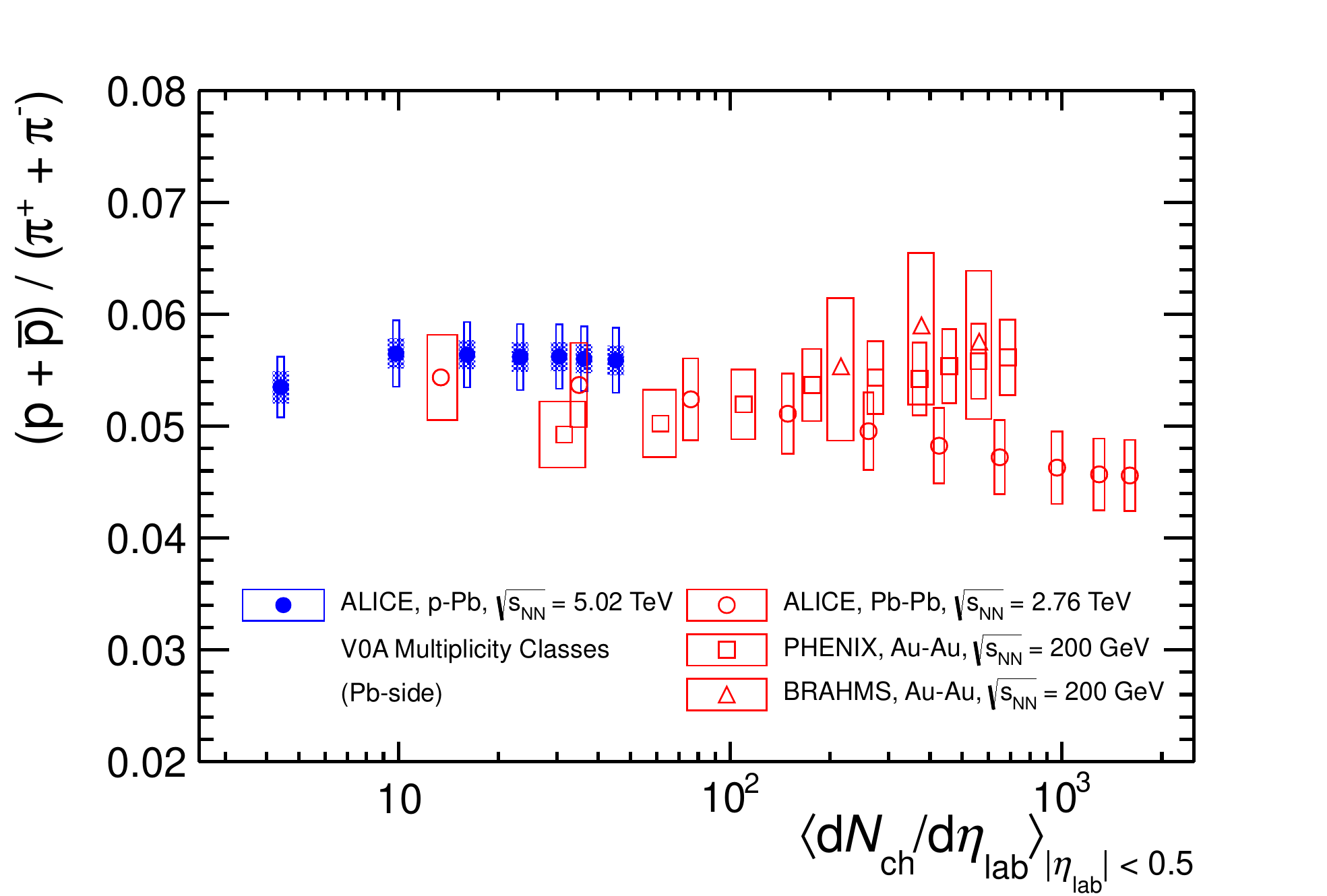}
  \includegraphics[width=0.55\textwidth]{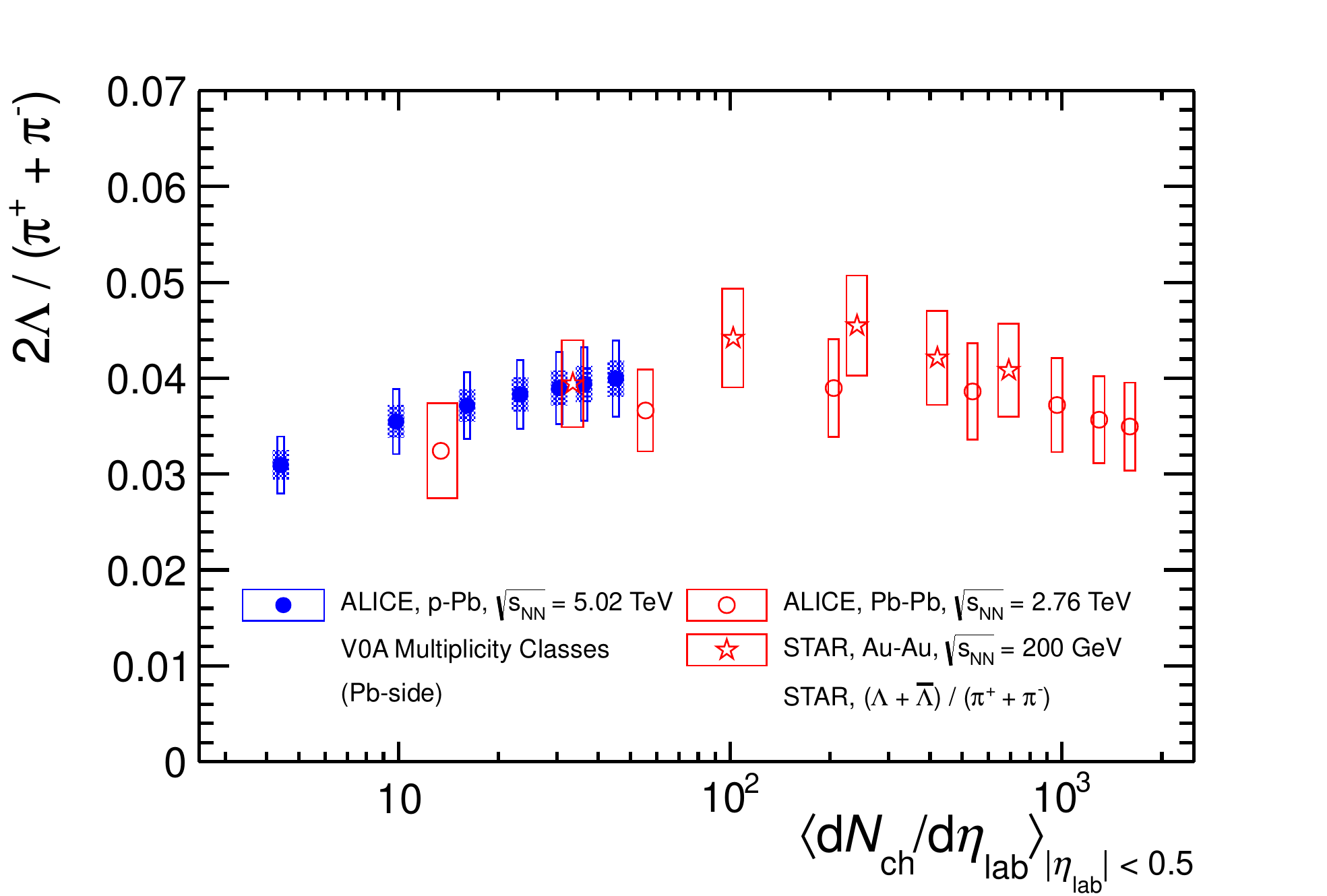}
\else
\begin{figure}[t!]
  \centering
  \includegraphics[width=0.5\textwidth]{pA_ka2pi_integrated_dNdeta}  
  \includegraphics[width=0.5\textwidth]{pA_pr2pi_integrated_dNdeta}
  \includegraphics[width=0.5\textwidth]{pA_lambda2pi_integrated_dNdeta}
\fi
  \caption{(color online) Particle yields d$N$/d$y$ of kaons, protons, and lambdas normalized to pions as a function of \dNdeta\ in each V0A multiplicity class (see text for details) measured in the rapidity interval $0 < \ycms < 0.5$. The values are compared to results obtained from \PbPb\ collisions at the LHC and Au--Au and d--Au collisions at RHIC measured at midrapidity. The empty boxes show the total systematic uncertainty; the shaded boxes indicate the contribution uncorrelated across multiplicity bins  (not estimated in Pb--Pb).}
  \label{fig:ratios}
\end{figure}

The \pt\ distributions of \allpart\ in $0 < \ycms\ < 0.5$ are shown 
in Fig.~\ref{fig:spectra} for different multiplicity intervals, 
as defined in Tab.~\ref{tab:multclasses}.
Particle/antiparticle as well as charged/neutral kaon transverse momentum distributions are identical within systematic uncertainties.

The \pt\ distributions show a clear evolution, becoming harder as the
multiplicity increases. The change is most pronounced for protons and
lambdas. They show an increase of the slope at low \pt, similar to the
one observed in heavy-ion collisions~\cite{prl-spectra,
  Abelev:2013vea}.  The stronger multiplicity dependence of the
spectral shapes of heavier particles is evident when looking at the ratios
\kpi~=~(\kap + \kam)/(\pip + \pim),  \ppi~=~(p + \pbar)/(\pip +
\pim) and  \lmb/\kzero as functions of
\pt, shown in Fig.~\ref{fig:RatiosVsPt} for the 0--5\% and 60--80\%
event classes.  The ratios \ppi\ and \lmb/\kzero\ show
a significant enhancement at intermediate \pt~$\sim 3$~\gevc,
qualitatively reminiscent of that measured in \PbPb\
collisions~\cite{prl-spectra, Abelev:2013vea, ALICE:2013xaa}.  The latter
are generally discussed in terms of collective flow or quark
recombination \cite{Fries:2003vb, Bozek:2011gq,Muller:2012zq}.
However, the magnitude of the observed effects differs significantly
between \pPb\ and in \PbPb. The maximum of the \ppi\ (\lmb/\kzero)
ratio reaches $\sim$ 0.8 (1.5) in central \PbPb\ collisions, but only
0.4 (0.8) in the highest multiplicity \pPb\ events.  The highest
multiplicity bin in \pPb\ collisions exhibits ratios of \ppi\ and
\lmb/\kzero\ which have maxima close to the corresponding ratios in
the 60-70\% bin in \PbPb\ collisions but differ somewhat in shape at
lower \pt.  The value of \dNdeta\ in central \pPb\ collisions (45 $\pm$ 1)
is a factor $\sim1.7$ lower than the one in the 60-70\%
\PbPb\ bin.  A similar enhancement of the \ppi\ ratio in
high-multiplicity d--Au collisions has also been reported for RHIC
energies~\cite{Adare:2013esx}.

It is worth noticing that the ratio \ppi\ as a function of \dNdeta\ in
a given \pt-bin follows a power-law behavior: $\frac{\rm p}{\pi}\left(\pt\right) = A(\pt)
\times \left[\dNdeta\right]^{B(\pt)}$. As shown in Fig.~\ref{fig:p-scaling} (top),
the same trend is also observed in \PbPb\ collisions. The exponent of
the power-law function exhibits the same value in both
collision systems (Fig.~\ref{fig:p-scaling}, middle). The same feature
is also observed in the \lmb/\kzero\ ratio (Fig.~\ref{fig:p-scaling},
bottom).

The \pt-integrated yields and \avpT\ are computed using the data in
the measured range and extrapolating them down to zero and to high
\pt\ (up to 10~\gevc). The fraction of extrapolated yield for high (low) multiplicity
events is about 8\% (9\%), 10\% (12\%), 7\% (13\%), 17\% (30\%) for
$\pi^{\pm}$, K$^{\pm}$, p and \pbar, \lmb\ and \almb\ respectively and is negligible for \kzero.
Several parametrizations have been tested, among which the blast-wave
function~\cite{Schnedermann:1993ws} (see below) gives the best
description of the data over the full \pt\ range
(Fig.~\ref{fig:spectra}). Other fit functions~\cite{Abelev:2008ez}
(Boltzmann, \mt-exponential, \pt-exponential, Tsallis-Levy,
Fermi-Dirac, Bose-Einstein) have been used to estimate the systematic
uncertainty on the extrapolation, restricting the range to low \pt\
for those functions not giving a satisfactory description of the data
over the full range. The uncertainty on the extrapolation amounts to
about 2\% for $\pi^{\pm}$, K$^{\pm}$, p(\pbar), 3\% (8\% in low
multiplicity events) for \lmb(\almb), and it is negligible for \kzero\
(since the \pt\ coverage ranges down to 0).

The \avpT\ increases with multiplicity, at a rate which is stronger
for heavier particles, as shown in Fig.~\ref{fig:meanpt}. A similar
mass ordering is also observed in \pp~\cite{Chatrchyan:2012qb} and
\PbPb~\cite{Abelev:2013vea} collisions as a function of multiplicity.

In Fig.~\ref{fig:ratios}, the ratios to the pion yields are compared to \PbPb\ results at the LHC and
Au--Au and d--Au results at
RHIC~\cite{Agakishiev:2011ar,Abelev:2008ez,Aggarwal:2010ig,Adare:2013esx,Adler:2003cb,Arsene:2005mr}. 
While the \ppi\ ratio shows no evolution from peripheral to central
events, a small increase is observed in the \kpi\ and \lpi\ ratios,
accounting for the bin-to-bin correlations of the uncertainties.  A
similar rise is observed in \PbPb, Au--Au and d--Au collisions.  This
is typically attributed to a reduced canonical suppression of
strangeness production in larger freeze-out
volumes~\cite{Sollfrank:1997bq} or to an enhanced strangeness
production in a quark-gluon plasma~\cite{Letessier:2006wn}.

The observations reported here are not strongly dependent on the
actual variable used to select multiplicity classes.  Alternative
approaches, such as using the total charge in both \VZEROA\ and
\VZEROC\ detectors, the energy deposited in the \ZNA\ 
(which originates from neutrons of the Pb nucleus)
and the number of clusters in the first ITS layers reveal very similar
trends. 
In the cases where the largest deviation is observed,
the \ppi\ ratio is essentially the same in
0-5\% events and it is $\sim$ 15\% higher at $\pt\sim3~\gevc$ in the
60-80\% class. Part of this difference is due to the mild correlation
of events at forward and central rapidity: the lowest multiplicity
class selected with \ZNA\ leads to a larger multiplicity at
midrapidity than the corresponding class selected with the \VZEROA.

\section{Discussion}

\ifplb
\begin{figure}[t]
  \centering
  \includegraphics[width=0.7\textwidth]{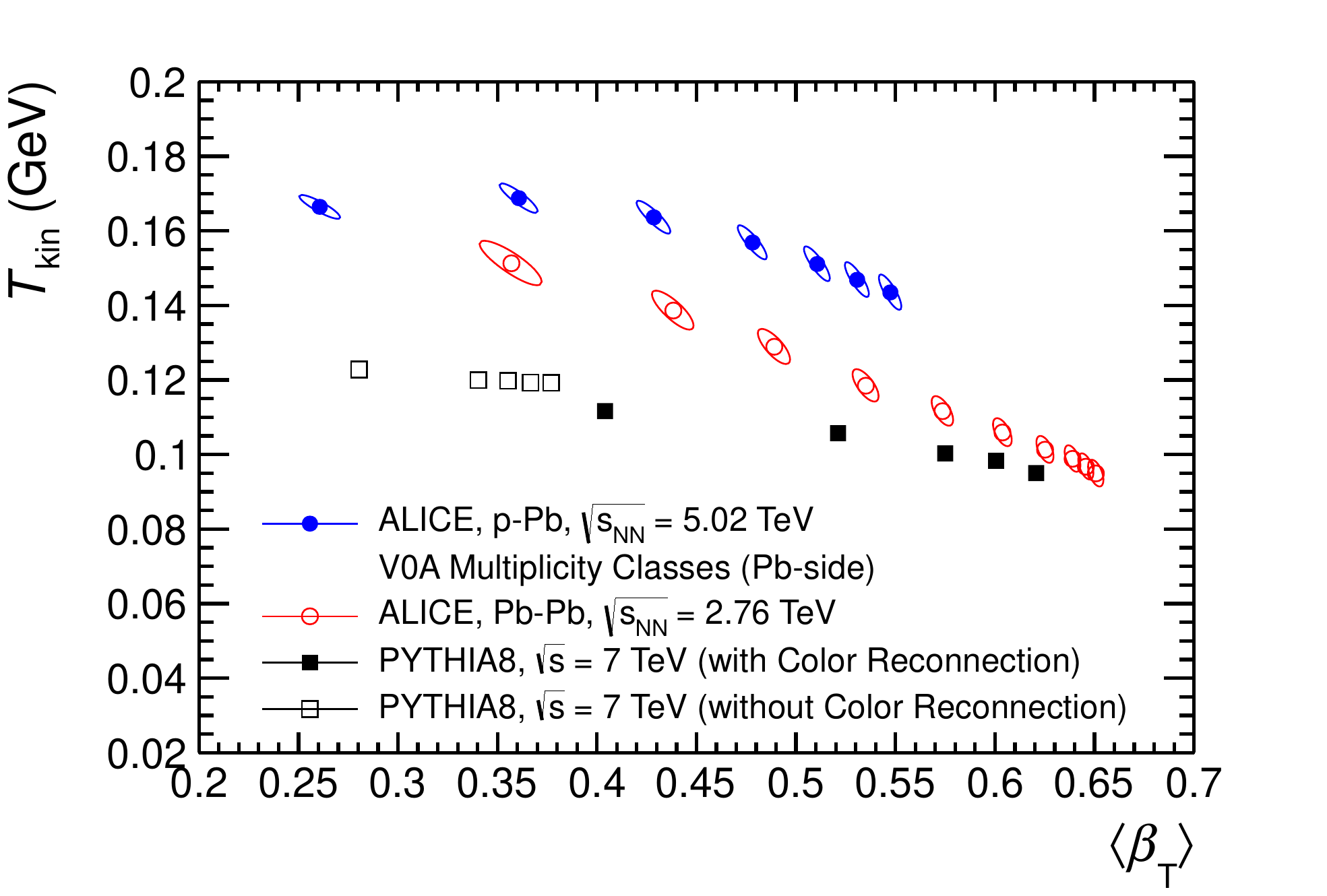}  
\else
\begin{figure}[t]
  \centering
  \includegraphics[width=0.5\textwidth]{pikapr_blastwave_V0A_withPYTHIA_cont}  
\fi
  \caption{(color online) Results of blast-wave fits, compared to \PbPb\ data and MC simulations from PYTHIA8 with and without color reconnection.
Charged-particle multiplicity increases from left to right.
Uncertainties from the global fit are shown as correlation ellipses.  }
  \label{fig:blast-wave}
\end{figure}

\ifplb
\begin{figure}[t!]
  \centering
  \includegraphics[trim=0cm 0cm 0cm 1.2cm, clip=true, width=0.7\textwidth]{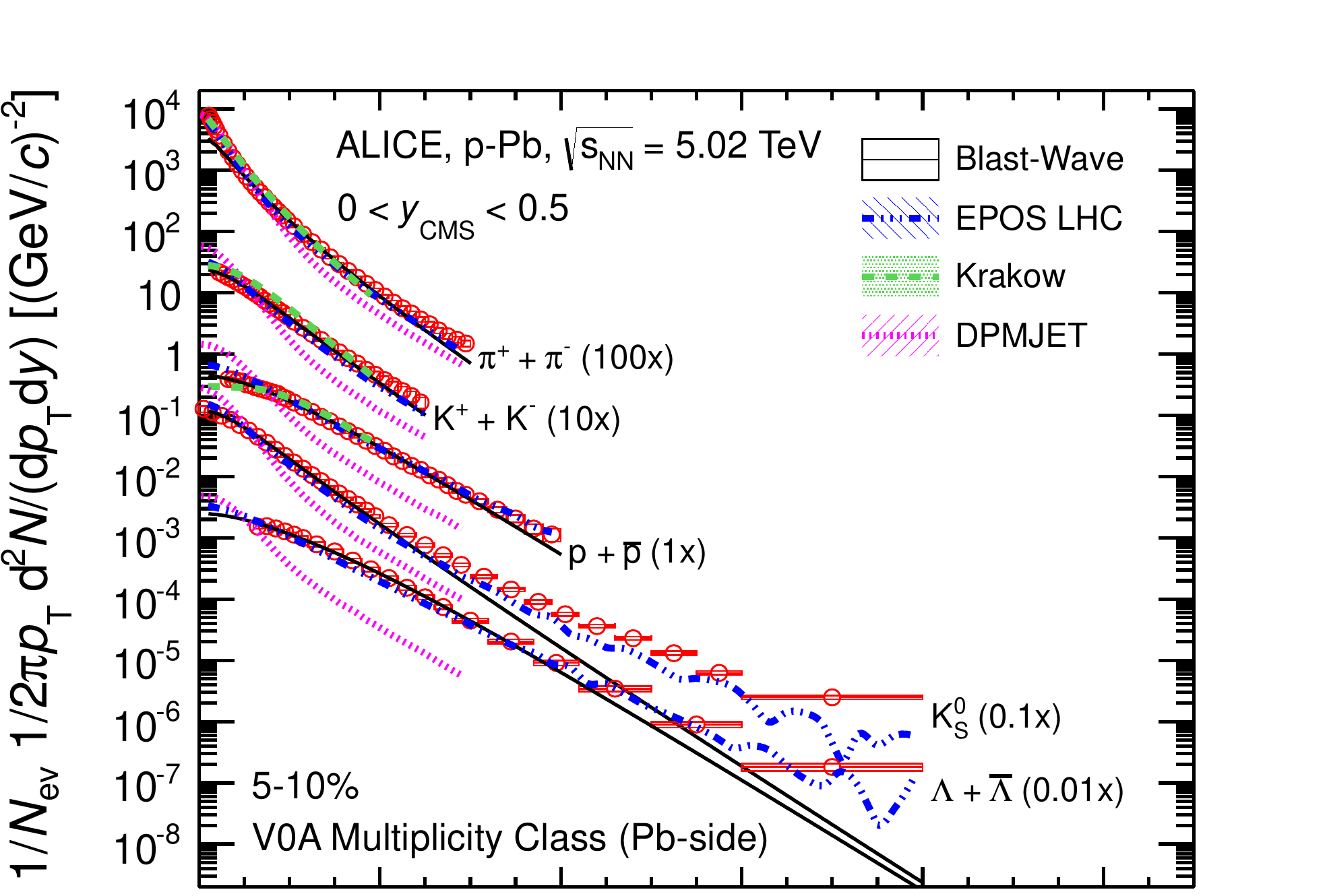}  

  \vspace{-0.08cm}

  \includegraphics[trim=0cm 0cm 0cm 1.2cm, clip=true, width=0.7\textwidth]{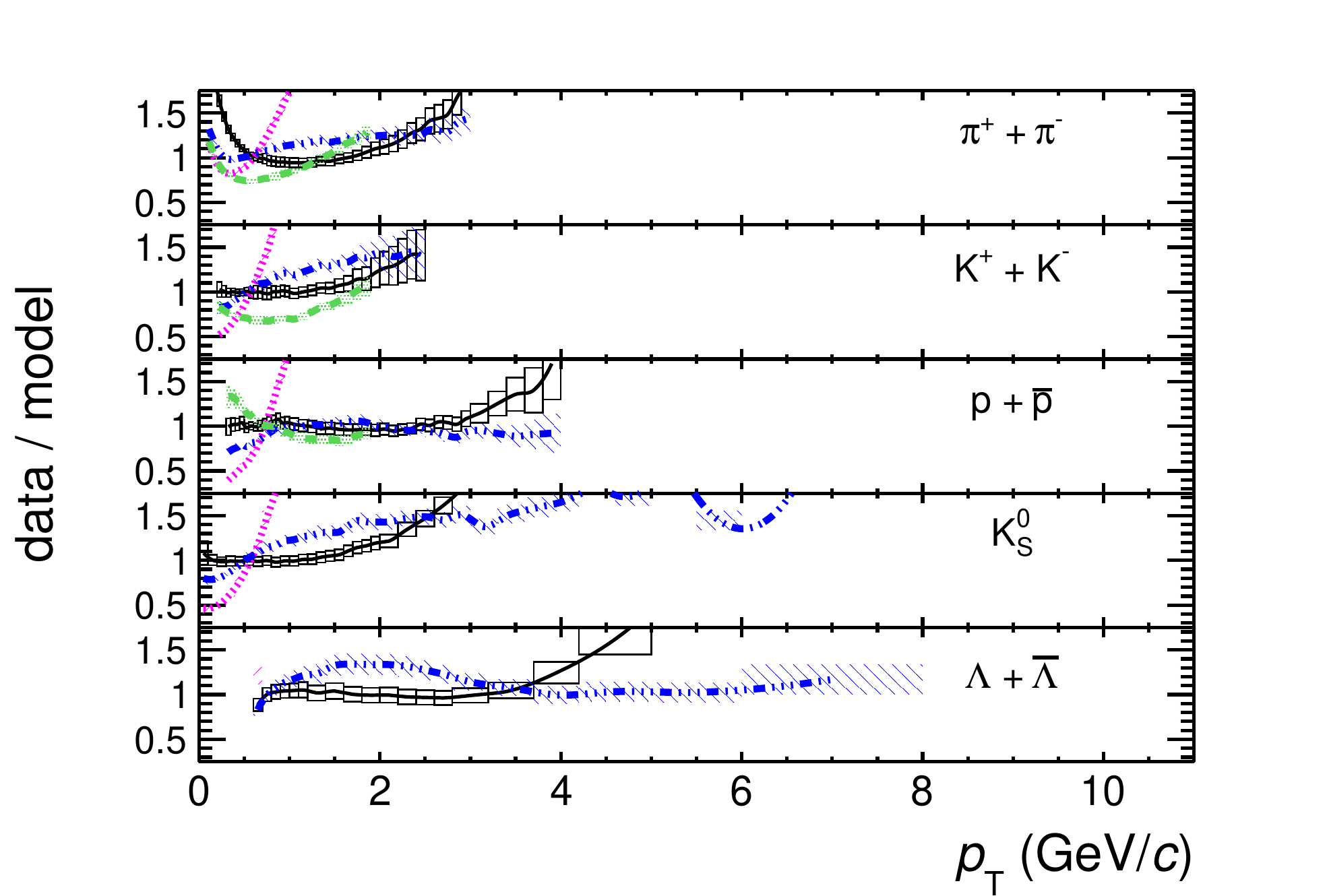}  
\else
\begin{figure}[t]
  \centering
  \includegraphics[trim=0cm 0cm 0cm 1.2cm, clip=true, width=0.5\textwidth]{BozekEpos}  

  \vspace{-0.08cm}

  \includegraphics[trim=0cm 0cm 0cm 1.2cm, clip=true, width=0.5\textwidth]{BozekEposRatio}  
\fi

  \caption{(color online) Pion, kaon, and proton transverse momentum distributions in the 5-10\% V0A multiplicity class measured in the rapidity interval $0 < \ycms < 0.5$ compared to the several models (see text for details).  }
  \label{fig:BozekComparison}
\end{figure}

In heavy-ion collisions, the flattening of transverse momentum distribution and its
mass ordering find their natural explanation in the collective radial
expansion of the system~\cite{Heinz:2004qz}. This picture can be
tested in a blast-wave framework with a simultaneous fit to all
particles for each multiplicity bin. This parameterization assumes a
locally thermalized medium, expanding collectively with a common
velocity field and undergoing an instantaneous common freeze-out.  The
blast-wave functional form is given by~\cite{Schnedermann:1993ws}

\begin{equation}
  \frac{1}{\pt} \frac{\mathrm{d}N}{\mathrm{d}\pt} \propto \int_0^R r \mathrm{d}r\, m_{\rm T}\, I_0 \left( \frac{p_{\rm T}\sinh \rho}{T_{kin}} \right) K_1 \left( \frac{m_{\rm T}\cosh \rho}{T_{kin}} \right),
  \label{eq:blast-wave}
\end{equation}

\noindent where the velocity profile $\rho$ is described by
\begin{equation}
 \rho = \tanh^{-1} \beta_{\rm T} = \tanh^{-1} \Biggl(\left(\frac{r}{R}\right)^{n} \beta_{s} \Biggr) \; .
 \label{eq:rhoBWdefintion}
\end{equation}

Here, $\mt = \sqrt{\pt^2+m^2}$ is the transverse mass, $I_0$ and $K_1$ 
are the modified Bessel functions, 
$r$ is the radial distance from the center of the fireball in
the transverse plane, $R$ is the radius of the fireball, $\beta_{\rm
  T}(r)$ is the transverse expansion velocity, $\beta_{s}$ is the
transverse expansion velocity at the surface, $n$ is the exponent of
the velocity profile and \Tfo\ is the kinetic freeze-out temperature.
The free parameters in the fit are \Tfo, $\beta_{s}$, $n$ and a
normalization parameter.

In contrast with the individual fits discussed above, the simultaneous
fit to all particle species under consideration can provide insight on the (common)
kinetic freeze-out properties of the system.  It has to be kept in
mind, however, that the actual values of the fit parameters depend substantially on
the fit range~\cite{Abelev:2013vea}.  In spite of this limitations, the blast-wave model still
provides a handy way to compare the transverse momentum distributions and their
evolution in different collision systems.

The fit presented in this Letter is performed in the same range as 
in~\cite{prl-spectra, Abelev:2013vea}, also including \kzero\
and \lmb(\almb). The ranges 0.5--1 \gevc, 0.2--1.5 \gevc, 0--1.5
\gevc, 0.3--3 \gevc\ and 0.6--3 \gevc\ have been used for \allpart\,
respectively. They have been defined according to the available data
at low \pt\ and based on the agreement with the data at high \pt,
justified considering that the assumptions underlying the
blast-wave model are not expected to be valid at high \pt.
Excluding the \kzero\ and \lmb(\almb) from the fit causes a
negligible difference in the fit parameters.

The results are reported in
Tab.~\ref{tab:blastwave_pA_default_withsys} and
Fig.~\ref{fig:blast-wave}.  Variations of the fit range lead to
large shifts ($\sim 10\%$) of the fit results (correlated across
centralities), as discussed for \PbPb\ data in~\cite{prl-spectra,
  Abelev:2013vea}

\begin{table*}[t]
  \centering
  \begin{tabular*}{\linewidth}{@{\extracolsep{\fill}}ccccc}
    \hline
    &&&\\[-0.7em]
    Event class & \avbT\ & \Tfo\ (\gevc) & n & $\chi^{2}$/ndf\\[0.3em]
    \hline
    &&&\\[-0.7em]
    0--5\% & 0.547 $\pm$ 0.006 $_{-0.02}^{+0.01}$ & 0.143 $\pm$ 0.005 $_{-0.01}^{+0.01}$ & 1.07 $\pm$ 0.03 $_{-0.09}^{+0.08}$ & 0.27 \\[0.3em]
    5--10\% & 0.531 $\pm$ 0.006 $_{-0.03}^{+0.01}$ & 0.147 $\pm$ 0.005 $_{-0.01}^{+0.01}$ & 1.14 $\pm$ 0.03 $_{-0.2}^{+0.1}$ & 0.33 \\[0.3em]
    10--20\% & 0.511 $\pm$ 0.007 $_{-0.03}^{+0.01}$ & 0.151 $\pm$ 0.005 $_{-0.01}^{+0.02}$ & 1.24 $\pm$ 0.04 $_{-0.2}^{+0.2}$ & 0.36 \\[0.3em]
    20--40\% & 0.478 $\pm$ 0.007 $_{-0.03}^{+0.02}$ & 0.157 $\pm$ 0.005 $_{-0.01}^{+0.02}$ & 1.41 $\pm$ 0.05 $_{-0.2}^{+0.2}$ & 0.35 \\[0.3em]
    40--60\% & 0.428 $\pm$ 0.009 $_{-0.03}^{+0.03}$ & 0.164 $\pm$ 0.004 $_{-0.02}^{+0.02}$ & 1.73 $\pm$ 0.07 $_{-0.4}^{+0.2}$ & 0.43 \\[0.3em]
    60--80\% & 0.36 $\pm$ 0.01 $_{-0.02}^{+0.04}$ & 0.169 $\pm$ 0.004 $_{-0.02}^{+0.02}$ & 2.4 $\pm$ 0.1 $_{-0.6}^{+0.2}$ & 0.54 \\[0.3em]
    80--100\% & 0.26 $\pm$ 0.01 $_{-0.01}^{+0.03}$ & 0.166 $\pm$ 0.003 $_{-0.01}^{+0.02}$ & 3.9 $\pm$ 0.3 $_{-0.7}^{+0.1}$ & 0.84 \\[0.3em]
    \hline
  \end{tabular*}
  \caption{Blast-wave parameters for simultaneous p--Pb fit of \allpart\ in the fit ranges  0.5--1 \gevc, 0.2--1.5 \gevc, 0--1.5
\gevc, 0.3--3 \gevc\ and 0.6--3 \gevc, respectively. Positive and negative variations of the parameters using the different fit ranges as done in~\cite{prl-spectra, Abelev:2013vea} are also reported.  }
  \label{tab:blastwave_pA_default_withsys}
\end{table*}

As can be seen in Fig.~\ref{fig:blast-wave}, the parameters show a
similar trend as the ones obtained in \PbPb. Within the limitations of
the blast-wave model, this observation is consistent with the presence
of radial flow in \pPb\ collisions.  A detailed comparison of the
resulting fit parameters between \PbPb~\cite{prl-spectra,
  Abelev:2013vea} and \pPb\
(Tab.~\ref{tab:blastwave_pA_default_withsys}) collisions shows that at
similar \dNdeta\ the values of parameters for \Tfo\ are similar for
the two systems, whereas the \avbT\ values are significantly higher in
\pPb\ collisions. While in \PbPb\ collisions high multiplicity events
are obtained through multiple soft interactions, in p--Pb collisions
the high multiplicity selection biases the sample towards harder
collisions~\cite{meanptpaper}. This could lead to the larger \avbT\
parameter obtained from the blast-wave fits. Under the assumptions of
a collective hydrodynamic expansion, a larger radial velocity in \pPb\
collisions has been suggested as a consequence of stronger radial
gradients in~\cite{Shuryak:2013ke}. 

In a hydrodynamically expanding system, the flow coefficients $v_n$
are also expected to exhibit a characteristic mass-dependent ordering
depending on the transverse expansion velocity. To probe this picture,
the \pt\ distributions are fitted simultaneously with the elliptic
flow coefficient extracted from two particle correlations $v_{2}$ of
\allpi, \allk, \allp\ measured in~\cite{pidcorr}, with the extension
of the blast-wave model of~\cite{Huovinen:2001cy}. This global fit is found to describe the
$v_{2}$ of pions, kaons and protons relatively well, even if the
quality of the fit is slightly worse than that of similar fits in \PbPb\
collisions, in particular for the proton $v_{2}$.  Compared to the
case where only the particle $\pt$-differential yields are used, the
fit results of \Tfo\ and \avbT\ differ by about 2\% only.

Other processes not related to hydrodynamic collectivity could also 
be responsible for the observed results.
This is illustrated in
Fig.~\ref{fig:blast-wave}, which shows the results obtained by
applying the same fitting procedure to transverse momentum distributions from the simulation of
pp collisions at \s~=~7 TeV with the PYTHIA8 event generator (tune
4C)~\cite{Corke:2010yf}, a model not including any collective system
expansion.  PYTHIA8 events are divided into several classes according to
the charged-particle multiplicity at midrapidity $\left| \hlab \right|
< 0.3$, namely $N_{\rm ch} < 5$, $5 \le N_{\rm ch} < 10$, $10 \le
N_{\rm ch} < 15$, $15 \le N_{\rm ch} < 20$ and $N_{\rm ch} \ge 20$.
The fit results are shown for PYTHIA8 simulations performed both with
and without the color reconnection
mechanism~\cite{Skands:2007zg,Schulz:2011qy}. This mechanism is
necessary in PYTHIA tunes to describe the evolution of \avpT\ with
multiplicity in pp collisions~\cite{meanptpaper}.  With color
reconnection the evolution of PYTHIA8 transverse momentum distributions follows a
similar trend as the one observed for p--Pb and Pb--Pb collisions at
the LHC, while without color reconnection it is not as strong.  This
generator study shows that other final state mechanisms, such as color
reconnection, can mimic the effects of radial flow~\cite{Ortiz:2013yxa}.

The \pt\ distributions in the 5-10\% bin are compared in
Fig.~\ref{fig:BozekComparison} with calculations from the DPMJET,
Krak\'ow~\cite{Bozek:2011if} and EPOS LHC 1.99
v3400~\cite{Pierog:2013ria} models.  The QCD-inspired
DPMJET~\cite{Roesler:2000he} generator, which is based on the
Gribov-Glauber approach, treats soft and hard scattering processes in
an unified way.  It has been found to successfully reproduce the
pseudorapidity distribution of charged particles in NSD p--Pb
collisions at the LHC as reported in~\cite{ALICE:2012xs}. On the other
hand, it cannot reproduce the \pt\ distribution~\cite{ALICE:2012mj} and
the \avpT\ of charged particles~\cite{meanptpaper}.  In the Krak\'ow
hydrodynamic model, fluctuating initial conditions are implemented based
on a Glauber model using a Monte Carlo simulation.  The expansion of
the system is calculated event-by-event in a 3+1 dimensional viscous
hydrodynamic approach and the freeze-out follows statistical
hadronization in a Cooper-Frye formalism. In the EPOS model, founded
on ``parton-based Gribov Regge theory'', the initial hard and soft
scattering creates ``flux tubes'' which either escape the medium and
hadronize as jets or contribute to the bulk matter, described in terms
of hydrodynamics.  The version of the model used here implements a
simplified treatment of the collective
expansion~\cite{Pierog:2013ria}. EPOS predictions including the full
hydrodynamic calculation~\cite{Werner:2012xh} are not available at the
time of writing.

The transverse momentum distributions in the 5-10\% multiplicity class
are compared to the predictions by Krak\'ow for $11 \leq N_{\rm part}
\leq 17$, since the \dNdeta\ from the model matches best with the
measured value in this class.  DPMJET and EPOS events have been
selected according to the charged particle multiplicity in the
\VZEROA\ acceptance in order to match the experimental selection.
DPMJET distributions are softer than the measured ones and the model
overpredicts the production of all particles for \pt\ lower than about
0.5--0.7 \gevc\ and underpredicts it at higher momenta.  At high-\pt,
the \pt\ spectra shapes of pions and kaons are rather well reproduced
for momenta above 1 and 1.5~\gevc\, respectively. Final state effects
may be needed in order to reproduce the data.  In fact, The Krak\'ow
model reproduces reasonably well the spectral shapes of pions and kaons below
transverse momenta of 1 \gevc\ where hydrodynamic effects are expected
to dominate. For higher momenta, the observed deviations for pions and
kaons could be explained in a hydrodynamic framework as due to the
onset of a non-thermal component. EPOS can reproduce the pion and
proton distributions within 20\% over the full measured range, while
larger deviations are seen for kaons and lambdas.  The yield and the
shape of the \pt\ distributions of protons are rather well described
by both models.  In contrast to a similar comparison for \PbPb\
collisions \cite{prl-spectra, Abelev:2013vea}, in the Krak\'ow
calculation the yield of pions and kaons seems to be overestimated. It
is interesting to notice that when final state interactions are
disabled in EPOS, the description of many \pp\ and \pPb\ observables
worsens significantly~\cite{Pierog:2013ria}. 

\section{Conclusions}

In summary, we presented a comprehensive measurement of \allpart\ in
\pPb\ collisions at \snn~=~5.02~TeV at the LHC. These data represent a
crucial set of constraints for the modeling of proton-lead collisions at the
LHC.  The transverse momentum distributions show a clear evolution
with multiplicity, similar to the pattern observed in high-energy pp and heavy-ion
collisions, where in the latter case the effect is usually attributed
to collective radial expansion.
Models incorporating final state effects give a better description of the data. 

 \newenvironment{acknowledgement}{\relax}{\relax}
 \begin{acknowledgement}
 \section{Acknowledgements}
We are grateful to P. Bozek, T. Pierog, and K. Werner for the useful
discussion and for providing the results of their calculations.

The ALICE collaboration would like to thank all its engineers and technicians for their invaluable contributions to the construction of the experiment and the CERN accelerator teams for the outstanding performance of the LHC complex.
\\
The ALICE collaboration acknowledges the following funding agencies for their support in building and
running the ALICE detector:
 \\
State Committee of Science,  World Federation of Scientists (WFS)
and Swiss Fonds Kidagan, Armenia,
 \\
Conselho Nacional de Desenvolvimento Cient\'{\i}fico e Tecnol\'{o}gico (CNPq), Financiadora de Estudos e Projetos (FINEP),
Funda\c{c}\~{a}o de Amparo \`{a} Pesquisa do Estado de S\~{a}o Paulo (FAPESP);
 \\
National Natural Science Foundation of China (NSFC), the Chinese Ministry of Education (CMOE)
and the Ministry of Science and Technology of China (MSTC);
 \\
Ministry of Education and Youth of the Czech Republic;
 \\
Danish Natural Science Research Council, the Carlsberg Foundation and the Danish National Research Foundation;
 \\
The European Research Council under the European Community's Seventh Framework Programme;
 \\
Helsinki Institute of Physics and the Academy of Finland;
 \\
French CNRS-IN2P3, the `Region Pays de Loire', `Region Alsace', `Region Auvergne' and CEA, France;
 \\
German BMBF and the Helmholtz Association;
\\
General Secretariat for Research and Technology, Ministry of
Development, Greece;
\\
Hungarian OTKA and National Office for Research and Technology (NKTH);
 \\
Department of Atomic Energy and Department of Science and Technology of the Government of India;
 \\
Istituto Nazionale di Fisica Nucleare (INFN) and Centro Fermi -
Museo Storico della Fisica e Centro Studi e Ricerche "Enrico
Fermi", Italy;
 \\
MEXT Grant-in-Aid for Specially Promoted Research, Ja\-pan;
 \\
Joint Institute for Nuclear Research, Dubna;
 \\
National Research Foundation of Korea (NRF);
 \\
CONACYT, DGAPA, M\'{e}xico, ALFA-EC and the EPLANET Program
(European Particle Physics Latin American Network)
 \\
Stichting voor Fundamenteel Onderzoek der Materie (FOM) and the Nederlandse Organisatie voor Wetenschappelijk Onderzoek (NWO), Netherlands;
 \\
Research Council of Norway (NFR);
 \\
Polish Ministry of Science and Higher Education;
 \\
National Authority for Scientific Research - NASR (Autoritatea Na\c{t}ional\u{a} pentru Cercetare \c{S}tiin\c{t}ific\u{a} - ANCS);
 \\
Ministry of Education and Science of Russian Federation, Russian
Academy of Sciences, Russian Federal Agency of Atomic Energy,
Russian Federal Agency for Science and Innovations and The Russian
Foundation for Basic Research;
 \\
Ministry of Education of Slovakia;
 \\
Department of Science and Technology, South Africa;
 \\
CIEMAT, EELA, Ministerio de Econom\'{i}a y Competitividad (MINECO) of Spain, Xunta de Galicia (Conseller\'{\i}a de Educaci\'{o}n),
CEA\-DEN, Cubaenerg\'{\i}a, Cuba, and IAEA (International Atomic Energy Agency);
 \\
Swedish Research Council (VR) and Knut $\&$ Alice Wallenberg
Foundation (KAW);
 \\
Ukraine Ministry of Education and Science;
 \\
United Kingdom Science and Technology Facilities Council (STFC);
 \\
The United States Department of Energy, the United States National
Science Foundation, the State of Texas, and the State of Ohio.

 \end{acknowledgement}
\bibliographystyle{h-physrev}

\newpage
\appendix
\section{The ALICE Collaboration}
\label{app:collab}



\begingroup
\small
\begin{flushleft}
B.~Abelev\Irefn{org69}\And
J.~Adam\Irefn{org36}\And
D.~Adamov\'{a}\Irefn{org77}\And
A.M.~Adare\Irefn{org126}\And
M.M.~Aggarwal\Irefn{org81}\And
G.~Aglieri~Rinella\Irefn{org33}\And
M.~Agnello\Irefn{org87}\textsuperscript{,}\Irefn{org104}\And
A.G.~Agocs\Irefn{org125}\And
A.~Agostinelli\Irefn{org25}\And
Z.~Ahammed\Irefn{org121}
% ...
\renewcommand\labelenumi{\textsuperscript{\theenumi}~}

\section*{Affiliation notes}
\renewcommand\theenumi{\roman{enumi}}
\begin{Authlist}
\item \Adef{0}Deceased
\item \Adef{idp3704080}{Also at: M.V.Lomonosov Moscow State University, D.V.Skobeltsyn Institute of Nuclear Physics, Moscow, Russia}
\item \Adef{idp3949120}{Also at: University of Belgrade, Faculty of Physics and "Vin\v{c}a" Institute of Nuclear Sciences, Belgrade, Serbia}
\item \Adef{idp4237872}{Permanent address: Konkuk University, Seoul, Korea}
\item \Adef{idp4763904}{Also at: Institute of Theoretical Physics, University of Wroclaw, Wroclaw, Poland}
\end{Authlist}

\section*{Collaboration Institutes}
\renewcommand\theenumi{\arabic{enumi}~}
\begin{Authlist}

\item \Idef{org1}A. I. Alikhanyan National Science Laboratory (Yerevan Physics Institute) Foundation, Yerevan, Armenia
\item \Idef{org2}Benem\'{e}rita Universidad Aut\'{o}noma de Puebla, Puebla, Mexico
\item \Idef{org3}Bogolyubov Institute for Theoretical Physics, Kiev, Ukraine
\item \Idef{org4}Budker Institute for Nuclear Physics, Novosibirsk, Russia
\item \Idef{org5}California Polytechnic State University, San Luis Obispo, California, United States
\item \Idef{org6}Central China Normal University, Wuhan, China
\item \Idef{org7}Centre de Calcul de l'IN2P3, Villeurbanne, France 
\item \Idef{org8}Centro de Aplicaciones Tecnol\'{o}gicas y Desarrollo Nuclear (CEADEN), Havana, Cuba
\item \Idef{org9}Centro de Investigaciones Energ\'{e}ticas Medioambientales y Tecnol\'{o}gicas (CIEMAT), Madrid, Spain
\item \Idef{org10}Centro de Investigaci\'{o}n y de Estudios Avanzados (CINVESTAV), Mexico City and M\'{e}rida, Mexico
\item \Idef{org11}Centro Fermi - Museo Storico della Fisica e Centro Studi e Ricerche ``Enrico Fermi'', Rome, Italy
\item \Idef{org12}Chicago State University, Chicago, United States
\item \Idef{org13}Commissariat \`{a} l'Energie Atomique, IRFU, Saclay, France
\item \Idef{org14}COMSATS Institute of Information Technology (CIIT), Islamabad, Pakistan
\item \Idef{org15}Departamento de F\'{\i}sica de Part\'{\i}culas and IGFAE, Universidad de Santiago de Compostela, Santiago de Compostela, Spain
\item \Idef{org16}Department of Physics Aligarh Muslim University, Aligarh, India
\item \Idef{org17}Department of Physics and Technology, University of Bergen, Bergen, Norway
\item \Idef{org18}Department of Physics, Ohio State University, Columbus, Ohio, United States
\item \Idef{org19}Department of Physics, Sejong University, Seoul, South Korea
\item \Idef{org20}Department of Physics, University of Oslo, Oslo, Norway
\item \Idef{org21}Dipartimento di Fisica dell'Universit\`{a} and Sezione INFN, Cagliari, Italy
\item \Idef{org22}Dipartimento di Fisica dell'Universit\`{a} and Sezione INFN, Trieste, Italy
\item \Idef{org23}Dipartimento di Fisica dell'Universit\`{a} and Sezione INFN, Turin, Italy
\item \Idef{org24}Dipartimento di Fisica dell'Universit\`{a} `La Sapienza` and Sezione INFN, Rome, Italy
\item \Idef{org25}Dipartimento di Fisica e Astronomia dell'Universit\`{a} and Sezione INFN, Bologna, Italy
\item \Idef{org26}Dipartimento di Fisica e Astronomia dell'Universit\`{a} and Sezione INFN, Catania, Italy
\item \Idef{org27}Dipartimento di Fisica e Astronomia dell'Universit\`{a} and Sezione INFN, Padova, Italy
\item \Idef{org28}Dipartimento di Fisica `E.R.~Caianiello' dell'Universit\`{a} and Gruppo Collegato INFN, Salerno, Italy
\item \Idef{org29}Dipartimento di Scienze e Innovazione Tecnologica dell'Universit\`{a} del Piemonte Orientale and Gruppo Collegato INFN, Alessandria, Italy
\item \Idef{org30}Dipartimento Interateneo di Fisica `M.~Merlin' and Sezione INFN, Bari, Italy
\item \Idef{org31}Division of Experimental High Energy Physics, University of Lund, Lund, Sweden
\item \Idef{org32}Eberhard Karls Universit\"{a}t T\"{u}bingen, T\"{u}bingen, Germany
\item \Idef{org33}European Organization for Nuclear Research (CERN), Geneva, Switzerland
\item \Idef{org34}Faculty of Engineering, Bergen University College, Bergen, Norway
\item \Idef{org35}Faculty of Mathematics, Physics and Informatics, Comenius University, Bratislava, Slovakia
\item \Idef{org36}Faculty of Nuclear Sciences and Physical Engineering, Czech Technical University in Prague, Prague, Czech Republic
\item \Idef{org37}Faculty of Science, P.J.~\v{S}af\'{a}rik University, Ko\v{s}ice, Slovakia
\item \Idef{org38}Frankfurt Institute for Advanced Studies, Johann Wolfgang Goethe-Universit\"{a}t Frankfurt, Frankfurt, Germany
\item \Idef{org39}Gangneung-Wonju National University, Gangneung, South Korea
\item \Idef{org40}Helsinki Institute of Physics (HIP), Helsinki, Finland
\item \Idef{org41}Hiroshima University, Hiroshima, Japan
\item \Idef{org42}Indian Institute of Technology Bombay (IIT), Mumbai, India
\item \Idef{org43}Indian Institute of Technology Indore, India (IITI)
\item \Idef{org44}Institut de Physique Nucl\'{e}aire d'Orsay (IPNO), Universit\'{e} Paris-Sud, CNRS-IN2P3, Orsay, France
\item \Idef{org45}Institut f\"{u}r Informatik, Johann Wolfgang Goethe-Universit\"{a}t Frankfurt, Frankfurt, Germany
\item \Idef{org46}Institut f\"{u}r Kernphysik, Johann Wolfgang Goethe-Universit\"{a}t Frankfurt, Frankfurt, Germany
\item \Idef{org47}Institut f\"{u}r Kernphysik, Technische Universit\"{a}t Darmstadt, Darmstadt, Germany
\item \Idef{org48}Institut f\"{u}r Kernphysik, Westf\"{a}lische Wilhelms-Universit\"{a}t M\"{u}nster, M\"{u}nster, Germany
\item \Idef{org49}Institut Pluridisciplinaire Hubert Curien (IPHC), Universit\'{e} de Strasbourg, CNRS-IN2P3, Strasbourg, France
\item \Idef{org50}Institute for High Energy Physics, Protvino, Russia
\item \Idef{org51}Institute for Nuclear Research, Academy of Sciences, Moscow, Russia
\item \Idef{org52}Institute for Subatomic Physics of Utrecht University, Utrecht, Netherlands
\item \Idef{org53}Institute for Theoretical and Experimental Physics, Moscow, Russia
\item \Idef{org54}Institute of Experimental Physics, Slovak Academy of Sciences, Ko\v{s}ice, Slovakia
\item \Idef{org55}Institute of Physics, Academy of Sciences of the Czech Republic, Prague, Czech Republic
\item \Idef{org56}Institute of Physics, Bhubaneswar, India
\item \Idef{org57}Institute of Space Science (ISS), Bucharest, Romania
\item \Idef{org58}Instituto de Ciencias Nucleares, Universidad Nacional Aut\'{o}noma de M\'{e}xico, Mexico City, Mexico
\item \Idef{org59}Instituto de F\'{\i}sica, Universidad Nacional Aut\'{o}noma de M\'{e}xico, Mexico City, Mexico
\item \Idef{org60}iThemba LABS, National Research Foundation, Somerset West, South Africa
\item \Idef{org61}Joint Institute for Nuclear Research (JINR), Dubna, Russia
\item \Idef{org62}Korea Institute of Science and Technology Information, Daejeon, South Korea
\item \Idef{org63}KTO Karatay University, Konya, Turkey
\item \Idef{org64}Laboratoire de Physique Corpusculaire (LPC), Clermont Universit\'{e}, Universit\'{e} Blaise Pascal, CNRS--IN2P3, Clermont-Ferrand, France
\item \Idef{org65}Laboratoire de Physique Subatomique et de Cosmologie (LPSC), Universit\'{e} Joseph Fourier, CNRS-IN2P3, Institut Polytechnique de Grenoble, Grenoble, France
\item \Idef{org66}Laboratori Nazionali di Frascati, INFN, Frascati, Italy
\item \Idef{org67}Laboratori Nazionali di Legnaro, INFN, Legnaro, Italy
\item \Idef{org68}Lawrence Berkeley National Laboratory, Berkeley, California, United States
\item \Idef{org69}Lawrence Livermore National Laboratory, Livermore, California, United States
\item \Idef{org70}Moscow Engineering Physics Institute, Moscow, Russia
\item \Idef{org71}National Centre for Nuclear Studies, Warsaw, Poland
\item \Idef{org72}National Institute for Physics and Nuclear Engineering, Bucharest, Romania
\item \Idef{org73}National Institute of Science Education and Research, Bhubaneswar, India
\item \Idef{org74}Niels Bohr Institute, University of Copenhagen, Copenhagen, Denmark
\item \Idef{org75}Nikhef, National Institute for Subatomic Physics, Amsterdam, Netherlands
\item \Idef{org76}Nuclear Physics Group, STFC Daresbury Laboratory, Daresbury, United Kingdom
\item \Idef{org77}Nuclear Physics Institute, Academy of Sciences of the Czech Republic, \v{R}e\v{z} u Prahy, Czech Republic
\item \Idef{org78}Oak Ridge National Laboratory, Oak Ridge, Tennessee, United States
\item \Idef{org79}Petersburg Nuclear Physics Institute, Gatchina, Russia
\item \Idef{org80}Physics Department, Creighton University, Omaha, Nebraska, United States
\item \Idef{org81}Physics Department, Panjab University, Chandigarh, India
\item \Idef{org82}Physics Department, University of Athens, Athens, Greece
\item \Idef{org83}Physics Department, University of Cape Town, Cape Town, South Africa
\item \Idef{org84}Physics Department, University of Jammu, Jammu, India
\item \Idef{org85}Physics Department, University of Rajasthan, Jaipur, India
\item \Idef{org86}Physikalisches Institut, Ruprecht-Karls-Universit\"{a}t Heidelberg, Heidelberg, Germany
\item \Idef{org87}Politecnico di Torino, Turin, Italy
\item \Idef{org88}Purdue University, West Lafayette, Indiana, United States
\item \Idef{org89}Pusan National University, Pusan, South Korea
\item \Idef{org90}Research Division and ExtreMe Matter Institute EMMI, GSI Helmholtzzentrum f\"ur Schwerionenforschung, Darmstadt, Germany
\item \Idef{org91}Rudjer Bo\v{s}kovi\'{c} Institute, Zagreb, Croatia
\item \Idef{org92}Russian Federal Nuclear Center (VNIIEF), Sarov, Russia
\item \Idef{org93}Russian Research Centre Kurchatov Institute, Moscow, Russia
\item \Idef{org94}Saha Institute of Nuclear Physics, Kolkata, India
\item \Idef{org95}School of Physics and Astronomy, University of Birmingham, Birmingham, United Kingdom
\item \Idef{org96}Secci\'{o}n F\'{\i}sica, Departamento de Ciencias, Pontificia Universidad Cat\'{o}lica del Per\'{u}, Lima, Peru
\item \Idef{org97}Sezione INFN, Bari, Italy
\item \Idef{org98}Sezione INFN, Bologna, Italy
\item \Idef{org99}Sezione INFN, Cagliari, Italy
\item \Idef{org100}Sezione INFN, Catania, Italy
\item \Idef{org101}Sezione INFN, Padova, Italy
\item \Idef{org102}Sezione INFN, Rome, Italy
\item \Idef{org103}Sezione INFN, Trieste, Italy
\item \Idef{org104}Sezione INFN, Turin, Italy
\item \Idef{org105}SUBATECH, Ecole des Mines de Nantes, Universit\'{e} de Nantes, CNRS-IN2P3, Nantes, France
\item \Idef{org106}Suranaree University of Technology, Nakhon Ratchasima, Thailand
\item \Idef{org107}Technical University of Split FESB, Split, Croatia
\item \Idef{org108}The Henryk Niewodniczanski Institute of Nuclear Physics, Polish Academy of Sciences, Cracow, Poland
\item \Idef{org109}The University of Texas at Austin, Physics Department, Austin, TX, United States
\item \Idef{org110}Universidad Aut\'{o}noma de Sinaloa, Culiac\'{a}n, Mexico
\item \Idef{org111}Universidade de S\~{a}o Paulo (USP), S\~{a}o Paulo, Brazil
\item \Idef{org112}Universidade Estadual de Campinas (UNICAMP), Campinas, Brazil
\item \Idef{org113}University of Houston, Houston, Texas, United States
\item \Idef{org114}University of Jyv\"{a}skyl\"{a}, Jyv\"{a}skyl\"{a}, Finland
\item \Idef{org115}University of Tennessee, Knoxville, Tennessee, United States
\item \Idef{org116}University of Tokyo, Tokyo, Japan
\item \Idef{org117}University of Tsukuba, Tsukuba, Japan
\item \Idef{org118}University of Zagreb, Zagreb, Croatia
\item \Idef{org119}Universit\'{e} de Lyon, Universit\'{e} Lyon 1, CNRS/IN2P3, IPN-Lyon, Villeurbanne, France
\item \Idef{org120}V.~Fock Institute for Physics, St. Petersburg State University, St. Petersburg, Russia
\item \Idef{org121}Variable Energy Cyclotron Centre, Kolkata, India
\item \Idef{org122}Vestfold University College, Tonsberg, Norway
\item \Idef{org123}Warsaw University of Technology, Warsaw, Poland
\item \Idef{org124}Wayne State University, Detroit, Michigan, United States
\item \Idef{org125}Wigner Research Centre for Physics, Hungarian Academy of Sciences, Budapest, Hungary
\item \Idef{org126}Yale University, New Haven, Connecticut, United States
\item \Idef{org127}Yonsei University, Seoul, South Korea
\item \Idef{org128}Zentrum f\"{u}r Technologietransfer und Telekommunikation (ZTT), Fachhochschule Worms, Worms, Germany
\end{Authlist}
\endgroup

\end{document}